\documentclass[pra,twocolumn,
superscriptaddress,
showpacs,
aps
]{revtex4-1}
\usepackage{graphicx,amsmath,amsfonts,amssymb,hyperref,color,dsfont,placeins,bm}

\newcommand{\beq}{\begin{equation}}
	\newcommand{\eeq}{\end{equation}}

\newcommand{\ope}{\hat{\mathcal{O}}}
\newcommand{\estate}{|e\rangle}

\newcommand{\lindblad}{\mathcal{L}}
\usepackage{kbordermatrix}
\renewcommand{\kbldelim}{(}
\renewcommand{\kbrdelim}{)}

\begin{document}
	\title{Superradiant emission of a thermal atomic beam into an optical cavity}
	\author{Simon~B.~J{\"a}ger}
	\affiliation{JILA, National Institute of Standards and Technology, and University of Colorado, Boulder, Colorado 80309-0440, USA}
	\author{Haonan~Liu}
	\affiliation{JILA, National Institute of Standards and Technology, and University of Colorado, Boulder, Colorado 80309-0440, USA}
	\author{John~Cooper} 
	\affiliation{JILA, National Institute of Standards and Technology, and University of Colorado, Boulder, Colorado 80309-0440, USA}
	\author{Travis~L.~Nicholson}
	\affiliation{Centre for Quantum Technologies, Department of Physics, National University of Singapore, Singapore 117543}
	\author{Murray~J.~Holland}
	\affiliation{JILA, National Institute of Standards and Technology, and University of Colorado, Boulder, Colorado 80309-0440, USA}
	
	\date{\today}
	
	\begin{abstract}
		We theoretically analyze the collective dynamics of a thermal beam of atomic dipoles that couple to a single mode when traversing an optical cavity. For this setup we derive a semiclassical model and determine the onset of superradiant emission and its stability. We derive analytical expressions for the linewidth of the emitted light and compare them with numerical simulations. In addition, we find and predict two different superradiant phases; a steady-state superradiant phase and a multi-component superradiant phase. In the latter case we observe sidebands in the frequency spectrum that can be calculated using a stability analysis of the amplitude mode of the collective dipole. We show that both superradiant phases are robust against free-space spontaneous emission and $T_2$ dephasing processes.
	\end{abstract}
	
	\maketitle
	
	\section{\label{sec: introduction}Introduction}
	The study of collective effects in atomic and molecular ensembles with cavity-mediated interactions is a very active research topic in quantum gas physics. Ongoing research focuses on the simulation and exploration of many-body systems~\cite{Baumann:2010, Ritsch:2013, Habibian:2013, Vaidya:2018, Muniz:2020} and also their application to metrology that takes advantage of the collective behavior~\cite{Schaeffer:2017, Swan:2018, Norcia:2018, Gothe:2019, Pedrozo:2020}. 
	
	An example of such a collective effect is superradiance, which describes the collective light emission enhanced by the build-up of macroscopic coherence in the ensemble of atomic or molecular dipoles. Originally, superradiance was predicted for free-space systems, that is, when the interparticle distance is smaller than the optical wavelength~\cite{Dicke:1954, Gross:1982}. However, this condition can be overcome by trapping the light in a confined volume, such as an optical cavity, and maintaining the condition of strong coupling of the particles to a single lossy resonator mode. More explicitly, superradiance in this case requires the cavity linewidth to be large compared to the collective linewidth of the dipoles. This results in a situation in which the coherence is stored in the atomic dipoles while the cavity mode is overdamped.
	
	The superradiant laser~\cite{Meiser:2009, Bohnet:2012} takes advantage of this effect and relies on a stable coherent collective dipole. This laser has the potential to produce light with an ultranarrow linewidth~\cite{Meiser:2009, Meiser:2010:1} that reflects the extremely high quality factor of the electronic transition~\cite{Norcia:2016:1, Norcia:2016:2}. In addition, recent studies have analyzed such systems as manifestation of phase synchronization~\cite{Xu:2014, Zhu:2015, Weiner:2017}, connected them to time crystals~\cite{Gong:2018, Iemini:2018, Tucker:2018, Barberena:2019, Booker:2020, Kessler:2020, Kessler:2021}, and discussed them as candidates for active optical clocks~\cite{Chen:2009, Zhang:2013}.
	
	A number of previous superradiant laser proposals and current experiments suggest trapping the atoms inside of the cavity ~\cite{Meiser:2009, Bohnet:2012, Meiser:2010:1, Meiser:2010:2, Maier:2014, Kraemer:2016, Debnath:2018, Zhang:2018, Laske:2019, Schaeffer:2020, Zhang:2021} with potential continuous incoherent repumping as its energy source. However, this is typically not easy to realize due to the need for closed transitions and external fields to trap the atoms. Furthermore, these additional complexities will usually lead to radiative heating of the atomic cloud and also to atom loss. 
	
	Another approach to achieve superradiant lasing is to couple a beam of moving atomic dipoles to a single resonator mode~\cite{Temnov:2005, Liu:2020, Jaeger:2021}. In this case the atoms can be precooled and prepared in the excited state before entering the cavity. This spatially separates the quantum state preparation stage from the collective emission that occurs while atoms travel through the cavity volume. Such designs are less prone to the adverse effects of radiative heating and may allow for an alternative pathway towards continuous-wave superradiant lasing in the optical domain~\cite{Liu:2020}.
	
	In this paper we study in detail the effect of Doppler broadening on collective emission when atoms traverse the optical resonator. We consider this to be the dominant broadening mechanism for metastable atomic dipoles and thermal atomic beams. We derive a general theoretical framework to study the collective emission of the atomic beam that includes a description of the atomic state when the atoms move through the cavity. This is then used to analyze the stability of the non-superradiant (NSR) and superradiant atomic configurations. For the latter, we predict a stable phase of the emitted light whereby phase diffusion is suppressed because of the formation of a large and robust collective dipole. Analyzing a realistic physical example, we show that superradiant emission is possible when the collective linewidth exceeds both the transit-time and Doppler broadening. In this regime we show that superradiant emission can appear in two forms; (i) steady-state superradiance (SSR), where the collective dipole is stable and phase diffusion dominates the dynamics of the collective dipole, and (ii) multi-component superradiance (MCSR), where the amplitude of the collective dipole oscillates in time. In the MCSR phase, we observe long-lived coherent oscillations in which the Doppler broadening itself is responsible for establishing the dynamical phase.
	
	This paper is organized as follows. In Sec.~\ref{sec: derivation} we introduce the model and derive the theoretical description that we will use throughout the paper. This description is analyzed in Sec.~\ref{sec: m-f} using a mean-field treatment. We derive the stability of the mean-field results and use them in Sec.~\ref{sec: analytical} to give analytical expressions for the linewidth of the emitted light. In Sec.~\ref{sec: thermal beam} we present the analysis of the dipole dynamics of a thermal beam traversing the cavity and compare simulation and analytical results. We conclude our discussion in Sec.~\ref{sec: conclusion}.
	
	\section{\label{sec: derivation}Derivation of the model}
	In this section we introduce the physical setup of the system and derive a theoretical description for it.
	
	\subsection{\label{subsec: master}System and master equation}
\begin{figure}[h!]
		\center
		\includegraphics[width=0.6\linewidth]{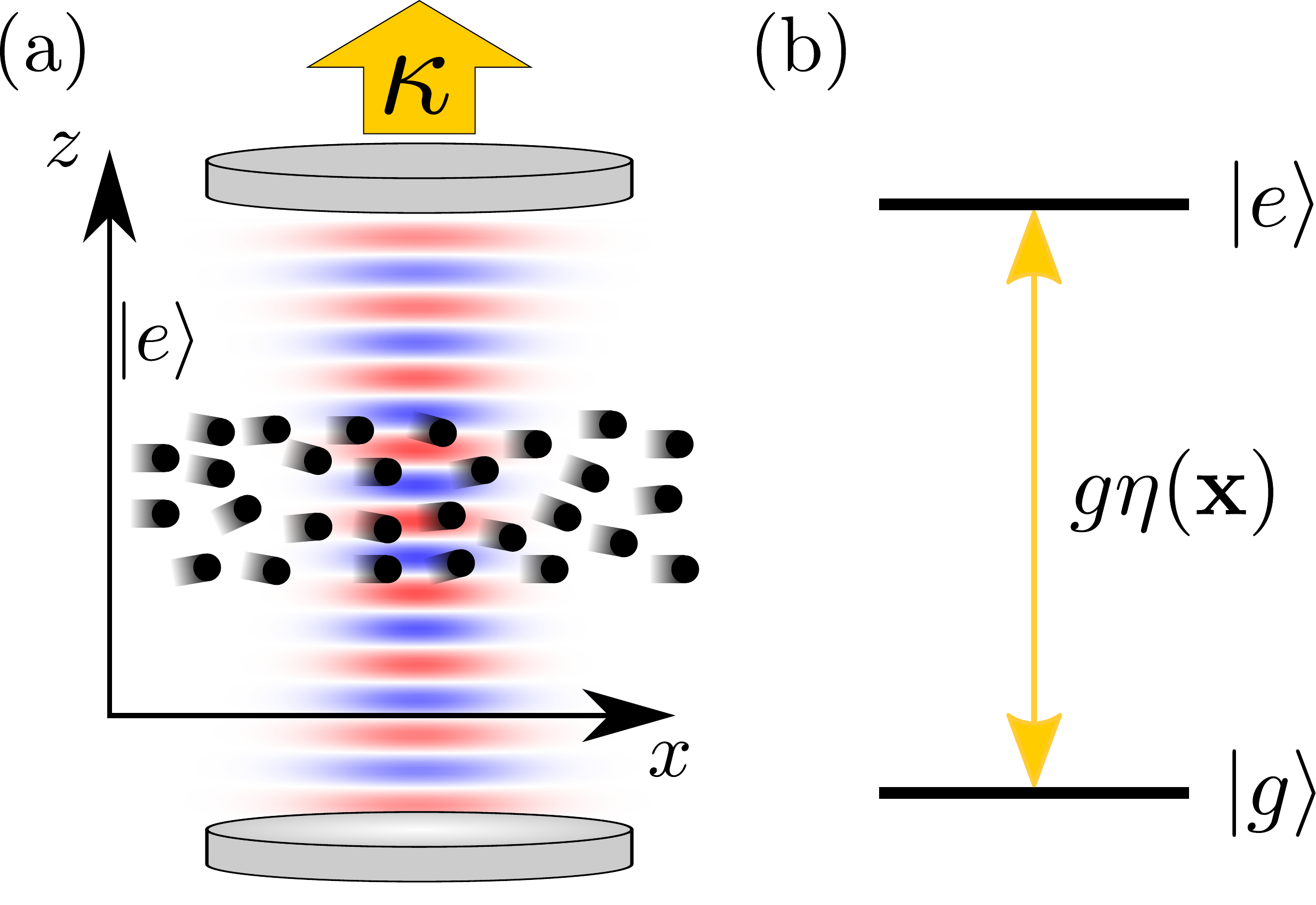}
		\caption{Schematic of the system (a) and the atom-cavity coupling (b). We consider a beam of two-level atoms in the excited state $|e\rangle$ traversing an optical cavity of loss rate $\kappa$ with a given velocity distribution. The $x$ and $z$ axes are chosen perpendicular and parallel to the cavity axis. The atomic beam is much broader than the optical wavelength $\lambda$ so that the atoms experience different phases of the cavity mode (blue and red denote different signs of the cavity mode function). The excited state $|e\rangle$ of the atomic dipoles (b) couples to the ground state $|g\rangle$ via photon emission into the cavity with coupling $g\eta({\bf x})$. The function $\eta({\bf x})$ is the mode function of the cavity.\label{Fig:1}}
	\end{figure}

	We consider a beam of metastable atomic dipoles with mass $m$ that travel through an optical cavity. Within the cavity the atoms couple to a single resonator mode. We choose $x$ and $z$ axes perpendicular and parallel to the cavity axis respectively [see Fig.~\ref{Fig:1}(a)].
	We describe the evolution of the atomic dipoles and the cavity field using a master equation for the density matrix $\hat{\rho}$, including internal and external degrees of freedom of the atoms and the cavity variables. The time evolution of $\hat{\rho}$ is given by
	\begin{align}
		\label{Mastereq}
		\frac{d \hat{\rho}}{d t}= \frac{1}{i\hbar}\left[\hat{H},\hat{\rho}\right]
		+ \kappa \lindblad[\hat{a}]\hat{\rho},
	\end{align}
	where $\lindblad[\ope]\hat{\rho} = \left(2\ope\hat{\rho}\ope^\dag - \ope^\dag\ope\hat{\rho} - \hat{\rho}\ope^\dag\ope\right)/2$ is the Lindblad superoperator.
	
	The first term in Eq.~\eqref{Mastereq} describes the coherent evolution and is governed by the Hamiltonian
	\begin{align}
		\label{H}
		\hat{H}=\sum_{j}\left[\frac{\hat{\bf p}_j^2}{2m}+\frac{\hbar g}{2}\eta(\hat{\bf x}_j)\left(\hat{a}^{\dag}\hat{\sigma}_j^-+\hat{\sigma}_j^{+}\hat{a}\right)\right],
	\end{align}
	which is presented in the frame rotating with the atomic transition frequency $\omega_a$. We have assumed the resonance condition of zero detuning between the cavity frequency $\omega_c$ and $\omega_a$, i.e., $\Delta_c \equiv \omega_c-\omega_a = 0$. The summation runs over all atoms in the beam. Inside the summation, the first term describes the atomic kinetic energy, and the second term describes the coherent coupling of atom $j$ to the single resonator mode. Here, ${\hat{\bf x}_j=(\hat{x}_j, \hat{y}_j, \hat{z}_j)^T}$ and ${\hat{\bf p}_j=(\hat{p}_{x,j}, \hat{p}_{y,j}, \hat{p}_{z,j})^T}$ are the position and momentum operators that satisfy the commutation relations $[\hat{\alpha}_j,\hat{p}_{\beta,k}]=i\hbar\delta_{jk}\delta_{\alpha\beta}$, with $\alpha,\beta\in\{x, y, z\}$. The function $g\eta({\bf \hat{x}})$ describes the coupling between the cavity and atoms [Fig.~\ref{Fig:1}(b)], where $g$ is the vacuum Rabi frequency at the field antinodes and $\eta({\bf x})$ is the spatial mode profile. The operators $\hat{a}$ and $\hat{a}^{\dag}$ are the photonic annihilation and creation operators that fulfill the usual bosonic commutation relation $[\hat{a},\hat{a}^{\dag}]=1$, while $\hat{\sigma}_j^+=|e\rangle_j\langle g|_j$ and $\hat{\sigma}_j^-=|g\rangle_j\langle e|_j$ are the atomic spin raising and lowering operators, where $|e\rangle_j$, $|g\rangle_j$ are the electronic excited and ground state of atom $j$, respectively.
	
	The second term in Eq.~\eqref{Mastereq} describes the leakage of cavity photons into the electromagnetic field modes external to the cavity. The rate $\kappa$ is the cavity decay rate and determines the linewidth of the cavity field mode when the atoms are not present. In the main part of this paper we will consider the cavity decay channel as the only source of decoherence, while we discuss additional noise sources in Sec.~\ref{subsec: spontaneous}.
	
	\subsection{Elimination of the cavity field}
	We describe our system in the superradiant regime where $\kappa$ exceeds all other atomic relaxation frequencies~\cite{Bonifacio:1971, Meiser:2010:1, Liu:2020}. In this regime we can adiabatically eliminate the fast cavity variables, which leads to an effective master equation for the atomic degrees of freedom described by the reduced density matrix
	\begin{align}
		\hat{\rho}_{\mathrm{atom}}=\mathrm{Tr}_{\mathrm{cav}}(\hat{\rho}),
	\end{align}
	where $\mathrm{Tr}_{\mathrm{cav}}(\,\ldots\,)$ denotes the partial trace over the cavity degrees of freedom. The resulting master equation for $\hat{\rho}_{\mathrm{atom}}$ reads
	\begin{align}
		\label{Mastereqat}
		\frac{d \hat{\rho}_{\mathrm{atom}}}{dt}=\frac{1}{i\hbar}\left[\sum_{j}\frac{\hat{\bf p}_j^2}{2m},\hat{\rho}_{\mathrm{atom}}\right]+\Gamma_c\mathcal{L}[\hat{J}^{-}]\hat{\rho}_\mathrm{atom},
	\end{align}
	where the incoherent part is governed by the single-atom linewidth
	\begin{align}
		\Gamma_c \equiv \frac{g^2}{\kappa}.
	\end{align}
	We have also introduced the generalized collective dipoles
	\begin{align}
		\hat{J}^{\pm}=\sum_j\eta(\hat{\bf x}_j)\hat{\sigma}_j^{\pm}.
	\end{align}
	
	For the remainder of the paper we focus on the dynamics of the atomic degrees of freedom. A useful description is given by the Heisenberg-Langevin equations that are equivalent to the master equation formalism, i.e.,
	\begin{align}
		\label{sigma-}
		\frac{d\hat{\sigma}_j^{-}}{dt}=&\frac{\Gamma_c}{2}\eta(\hat{\bf x}_j)\hat{\sigma}_j^{z}\hat{J}^{-}+\hat{\mathcal S}^{-}_j,\\
		\label{sigmaz}
		\frac{d\hat{\sigma}_j^{z}}{dt}=&-\Gamma_c\eta(\hat{\bf x}_j)\left(\hat{J}^{+}\hat{\sigma}_j^{-}+\hat{\sigma}_j^{+}\hat{J}^{-}\right)+\hat{\mathcal S}^{z}_j,\\
		\label{hatx}
		\frac{d\hat{\bf x}_j}{dt}=&\frac{\hat{\bf p}_j}{m},\\
		\label{hatp}
		\frac{d\hat{\bf p}_j}{dt}=&\frac{i\hbar\Gamma_c}{2}(\hat{\sigma}_j^{+}\hat{J}^{-}-\hat{J}^{+}\hat{\sigma}_j^{-})\left.\nabla_{\bf x}\eta({\bf x})\right|_{{\bf x}=\hat{\bf x}_j}+\hat{\mathcal{N}}_j,
	\end{align}
	where the noise terms are given by ${\hat{\mathcal{S}}_j^{-}=\eta(\hat{\bf x}_j)\hat{\sigma}^z_j\hat{\mathcal{F}}^{-}}$, ${\hat{\mathcal{S}}_j^{z}=-2\eta(\hat{\bf x}_j)(\hat{\mathcal{F}}^{+}\hat{\sigma}^{-}_j+\hat{\sigma}^{+}_j\hat{\mathcal{F}}^{-})}$ for internal degrees of freedom and by ${\hat{\mathcal{N}}_j=i\hbar\left.\nabla_{\bf x}\eta({\bf x})\right|_{{\bf x}=\hat{\bf x}_j}(\hat{\sigma}^{+}_j\hat{\mathcal{F}}^{-}-\hat{\mathcal{F}}^{+}\hat{\sigma}^{-}_j)}$ for the external force acting on atom $j$. The terms $\hat{\mathcal{F}}^{\pm}$ are effective noise terms on the coarse-grained timescale on which this system of equations evolve and satisfy the correlations $\langle\hat{\mathcal{F}}^{-}(t)\hat{\mathcal{F}}^{-}(t')\rangle_q=0=\langle\hat{\mathcal{F}}^{+}(t)\hat{\mathcal{F}}^{-}(t')\rangle_q$ and $\langle\hat{\mathcal{F}}^{-}(t)\hat{\mathcal{F}}^{+}(t')\rangle_q=\Gamma_c\delta(t-t')$, $\hat{\mathcal{F}}^{+}=(\hat{\mathcal{F}}^{-})^{\dag}$. The expectation value $\langle\,.\,\rangle_q$ is over the cavity degrees of freedom and the free-space photonic modes external to the cavity.
	
	\subsection{\label{sec: further approx}Parameter regime and $c$-number approximations}
	Our theoretical description is used to analyze the dynamics of the atoms that travel ballistically through the cavity. This requires neglecting optomechanical forces in Eq.~\eqref{hatp} by assuming
	\begin{equation}
		\label{hatp ballistic}
		\frac{d\hat{\bf p}_j}{dt}=0
	\end{equation}
	for all atoms. We discuss the validity of this approximation in Appendix~\ref{App:OptoForces}. Moreover, we will mostly work in the regime where atoms collectively emit into the cavity mode. This is possible if the transit time $\tau$ of an individual atom is of the same order of magnitude as the characteristic timescale of superradiant emission $1/(N\Gamma_c)$, where $N$ is the mean intracavity atom number.
	
	In order to simulate the Heisenberg-Langevin equations in Eqs.~\eqref{sigma-}--\eqref{hatp}, we make a semiclassical approximation where we exchange the quantum operators by $c$-numbers and use noise terms that simulate quantum noise~\cite{Schachenmayer:2015, Liu:2020, Jaeger:2021}. This semiclassical description can be derived by first writing down the Heisenberg-Langevin equations for the dipole components $\hat{\sigma}^x_j=\hat{\sigma}^{-}_j+\hat{\sigma}^{+}_j$, $\hat{\sigma}^y_j=i(\hat{\sigma}^{-}_j-\hat{\sigma}^{+}_j)$, $\hat{\sigma}^z$ and then exchanging them with their corresponding $c$-number equivalents $s_j^{x}$, $s_j^{y}$, and $s_j^{z}$. The same approach is repeated with the external operators $\hat{\bf x}_j$ and $\hat{\bf p}_j$ that are replaced by their corresponding classical counterparts ${\bf x}_j$ and ${\bf p}_j$. With this procedure we obtain the following $c$-number stochastic differential equations
	\begin{align}
		\label{sx}
		\frac{ds_j^{x}}{dt}=&\frac{\Gamma_c}{2}\eta({\bf x}_j)s_j^{z}J^{x}+\mathcal{S}^{x}_j,\\
		\label{sy}
		\frac{ds_j^{y}}{dt}=&\frac{\Gamma_c}{2}\eta({\bf x}_j)s_j^{z}J^{y}+\mathcal{S}^{y}_j,\\
		\label{sz}
		\frac{ds_j^{z}}{dt}=&-\frac{\Gamma_c}{2}\eta({\bf x}_j)\left(J^{x}s_j^{x}+J^{y}s_j^{y}\right)+\mathcal{S}^{z}_j,\\
		\label{x}
		\frac{d{\bf x}_j}{dt}=&\frac{{\bf p}_j}{m},
	\end{align}
	where 
	\begin{align}
		\label{Jaold}
		J^\alpha=\sum_j\eta({\bf x}_j)s_j^{\alpha},\,\,\,\,\alpha\in\{x,y\}.
	\end{align}
	are the $c$-number collective dipole components. We have neglected single-atom terms in Eqs.~\eqref{sx}--\eqref{x} that scale with $\Gamma_c$ compared to the collective terms that scale with $N\Gamma_c$. The noise terms are defined by $\mathcal{S}^{\alpha}_j=\eta({\bf x}_j)s^z_j\mathcal{F}^{\alpha}$, $\alpha\in\{x,y\}$ and $\mathcal{S}^{z}_j=-\eta({\bf x}_j)(s^x_j\mathcal{F}^{x}+s^y_j\mathcal{F}^{y})$. The independent random noise terms $\mathcal{F}^{x}$ and $\mathcal{F}^{y}$ fulfill $\langle \mathcal{F}^{x}(t)\mathcal{F}^{x}(t')\rangle=\Gamma_c\delta(t-t')=\langle \mathcal{F}^{y}(t)\mathcal{F}^{y}(t')\rangle$. These equations have been derived using the symmetric orderings of the operators and replacing these by their classical $c$-number counterparts~\cite{Liu:2020}.
	
	Beside the noise that is induced by $\mathcal{F}^{x}$ and $\mathcal{F}^{y}$ we also need to include another noise source that arises from introducing new atoms into the cavity. We assume throughout this paper that the atoms enter in the excited state $|e\rangle$. In that case an atom indexed by $j$ enters the cavity with $s_j^z=1$. Since the atom is in $|e\rangle$, the quantum uncertainty in $s_j^x$ and $s_j^y$ is maximal. This is modeled by randomly and independently initializing $s_j^x=\pm1$ and $s_j^y=\pm 1$~\cite{Schachenmayer:2015}. With this methodology we fulfill up to second order the correct initial spin-moments for the entering atoms, i.e., $\langle s^\alpha_j \rangle = \langle \hat{\sigma}^{\alpha}_j \rangle$, $\langle s_j^\alpha s_k^\alpha \rangle=\langle \hat{\sigma}_j^\alpha\hat{\sigma}_k^\alpha\rangle=\delta_{jk}$,  $\alpha\in\{x,y\}$, and $\langle s_j^xs_k^y \rangle = \langle \{\hat{\sigma}_j^x\hat{\sigma}_k^y\}_{\mathrm{sym}} \rangle = 0$, where $\delta_{jk}$ is Kronecker-delta and $\{\hat{\sigma}_j^x\hat{\sigma}_k^y\}_{\mathrm{sym}}\equiv \left(\hat{\sigma}_j^x\hat{\sigma}_k^y+\hat{\sigma}_k^y\hat{\sigma}_j^x\right)/2$ is the symmetric ordering of operators $\hat{\sigma}_j^x$ and $\hat{\sigma}_k^y$.
	
	In the next subsection we will apply Eqs.~\eqref{sx}--\eqref{x} with the noise terms introduced above to derive a phase-space density description of the atomic dipoles.
	
	\subsection{Phase-space density description}
	The phase-space density description of our model is derived by defining the classical phase-space density and the spin densities of the atomic beam as
	\begin{align}
		f({\bf x},{\bf p},t)=&\sum_{j}\delta({\bf x}-{\bf x}_j)\delta({\bf p}-{\bf p}_j),\\
		s^{\alpha}({\bf x},{\bf p},t)=&\sum_{j}s_j^{\alpha}\delta({\bf x}-{\bf x}_j)\delta({\bf p}-{\bf p}_j),
	\end{align}
	where $s_j^{\alpha}$ is the single-atom spin component with $\alpha\in\{x,y,z\}$. The collective dipole components defined in Eq.~\eqref{Jaold} are given by
	\begin{align}
		\label{Janew}
		J^\alpha=\int d{\bf x}\int d{\bf p}\,\eta({\bf x})s^\alpha({\bf x},{\bf p},t),\,\,\,\,\alpha\in\{x,y\},
	\end{align}
	and Eqs.~\eqref{sx}--\eqref{x} can be rewritten with density variables as
	\begin{align}
		\label{f}  
		\frac{\partial f}{\partial t}+\frac{\bf p}{m}\cdot \nabla_{\bf x}f=&0,\\
		\label{sxdensity}
		\frac{\partial s^x}{\partial t}+\frac{\bf p}{m}\cdot \nabla_{\bf x}s^x=&\frac{\Gamma_c}{2}\eta({\bf x})s^zJ^x+\mathcal{S}^x,\\
		\label{sydensity}
		\frac{\partial s^y}{\partial t}+\frac{\bf p}{m}\cdot \nabla_{\bf x}s^y=&\frac{\Gamma_c}{2}\eta({\bf x})s^zJ^y+\mathcal{S}^y,\\
		\label{szdensity}
		\frac{\partial s^z}{\partial t}+\frac{\bf p}{m}\cdot \nabla_{\bf x}s^z=&-\frac{\Gamma_c}{2}\eta({\bf x})\left(J^{x}s^{x}+J^{y}s^{y}\right)+\mathcal{S}^{z}.
	\end{align}
	Here, Eq.~\eqref{f} describes the free flight of the atomic beam. The noise terms are given by $\mathcal{S}^\alpha=\eta({\bf x})\mathcal{F}^\alpha s^z$ and $\mathcal{S}^z=-\eta({\bf x})\left(\mathcal{F}^xs^x+\mathcal{F}^ys^y\right)$. We emphasize that these noise terms are still local in time but long range in space.
	
	The initial conditions for the atoms entering the cavity can be formulated as noisy spatial boundary conditions for the stochastic partial differential equations~\eqref{f}--\eqref{szdensity}. In order to formulate these boundary conditions, we define $x=-x_0$ as the position on $x$ axis where the atoms enter the cavity. Notice that the exact choice of $x_0$ depends on the choice of the mode function $\eta({\bf x})$ and can in principle be $x_0=\infty$. We assign
	\begin{align}
		f(-x_0,y,z,{\bf p},t)=&f_0(y,z,{\bf p},t),\\
		s^x(-x_0,y,z,{\bf p},t)=&W^x(y,z,{\bf p},t),\\
		s^y(-x_0,y,z,{\bf p},t)=&W^y(y,z,{\bf p},t),\\
		s^z(-x_0,y,z,{\bf p},t)=&f_0(y,z,{\bf p},t)
	\end{align}
	as the initial conditions for the system at every instant of time $t$. Here, we have used
	\begin{align}
		f_0(y,z,{\bf p},t)=\sum_j\delta({\bf x}_0-{\bf x}_j)\delta({\bf p}-{\bf p}_j),
	\end{align}
	and ascribed ${{\bf x}_0=(-x_0,y,z)^T}$ to be the entrance surface. Since the atoms enter the cavity in $\estate$, the boundary conditions for $f$ and $s^z$ are the same. The initial noise terms in the $s^x$ and $s^y$ components can be described by
	\begin{align}
		W^\alpha(y,z,{\bf p},t)=\sum_js_j^\alpha\delta({\bf x}_0-{\bf x}_j)\delta({\bf p}-{\bf p}_j),\,\,\,\,\alpha\in\{x,y\}.
	\end{align}
	These noise terms have the second moments
	\begin{align}
		\label{noise}
		\langle W^{\alpha}(W^{\beta})'\rangle = &\frac{m}{p_x}\delta_{\alpha\beta}\delta(t-t')\delta(y-y')\delta(z-z') \notag\\
		&\times \delta({\bf p}-{\bf p}')f_0(y,z,{\bf p},t),
	\end{align}
	where we have simplified notation as $W^{\alpha}=W^{\alpha}(y,z,{\bf p},t)$ and $(W^{\beta})'=W^{\beta}(y',z',{\bf p}',t')$. Notice that such noise processes are both spatially and temporally local.
	
	Throughout this paper we will assume that the distribution of the atoms is spatially homogeneous. This requires that the diameter of the atomic beam is much larger than $\lambda$ [see Fig.~\ref{Fig:1}(a)] and the cavity waist $w$. This assumption allows for the formulation of an averaged atomic density $\rho({\bf p})$ using the ensemble average $\langle\,.\,\rangle_{\mathrm{ens}}$ of the boundary condition $f_0(y,z,{\bf p},t)$, i.e.,
	\begin{align}
		\rho({\bf p}) \equiv \langle f_0(y,z,{\bf p},t)\rangle_{\mathrm{ens}},
	\end{align}
	which is independent of space and time. As a result, after a time $t$ that is much larger than $\tau$, we achieve a stationary state for $f$ that satisfies $\langle f\rangle_{\mathrm{ens}}=\rho({\bf p})$ and describes a spatially homogeneous atomic density in the cavity mode volume. However, this does not imply that the spin densities $s^{a}$ are spatially homogeneous, which can already be seen in a mean-field description.
	
	\section{\label{sec: m-f}Mean-field analysis}
	In order to describe the mean-field dynamics of the spin densities, we discard for the moment any noise terms introduced by $W^{\alpha}$ and $\mathcal{F}^{\alpha}$, $\alpha\in\{x,y\}$. The resulting partial differential equations from Eqs.~\eqref{sxdensity}--\eqref{szdensity} read
	\begin{align}
		\label{meanfield sx}
		\frac{\partial s^x}{\partial t}+\frac{\bf p}{m}\cdot\nabla_{\bf x}s^x=&\frac{\Gamma_c}{2}\eta({\bf x})J^xs^z,\\
		\label{meanfield sy}
		\frac{\partial s^y}{\partial t}+\frac{\bf p}{m}\cdot\nabla_{\bf x}s^y=&\frac{\Gamma_c}{2}\eta({\bf x})J^ys^z,\\
		\label{meanfield sz}
		\frac{\partial s^z}{\partial t}+\frac{\bf p}{m}\cdot\nabla_{\bf x}s^{z}=&-\frac{\Gamma_c}{2}\eta({\bf x})\left(J^xs^x+J^ys^y\right).
	\end{align}
	In the following two subsections we will distinguish between the case when there is no superradiance $J^x = J^y = 0$ and when there is superradiance $(J^x,J^y)\neq (0,0)$.
	
	\subsection{\label{subsec: non-sr}Non-superradiant phase (NSR)}
	The system is in the non-superradiant phase (NSR) when there is no collective dipole, i.e., $J^x = J^y = 0$. In this phase, the mean-field stationary state is given by
	\begin{align}
		\label{sx_nsr}
		s^x=&0,\\
		s^y=&0,\\
		\label{sz_nsr}
		s^z=&\rho({\bf p}).
	\end{align}
	Here, we only report the density inside of the cavity for $t\gg\tau$.
	
	Although Eqs.~\eqref{sx_nsr}--\eqref{sz_nsr} always represent a stationary solution of the mean-field equations, they are not necessarily stable. Any noise, for instance introduced by $W^{\alpha}$ and $\mathcal{F}^{\alpha}$, could potentially destabilize the stationary state. 
	
	In order to determine the stability of the NSR phase, we calculate the evolution of small fluctuations in spin densities by letting $s^x=\delta s^x$ and $s^y=\delta s^y$ and $s^z=\rho({\bf p})+\delta s^z$. We do not need to specify the source of these small terms explicitly, but note that such fluctuations will be introduced by the noise processes when extending the theory to the full description of the dipole densities.
	
	The equations for $\delta s^x$, $\delta s^y$, and $\delta s^z$ are given by
	\begin{align}
		\label{deltasx}
		\frac{\partial \delta s^x}{\partial t}+\frac{\bf p}{m}\cdot\nabla_{\bf x}\delta s^{x}\approx&\frac{\Gamma_c}{2}\eta({\bf x})\delta J^x\rho({\bf p}),\\
		\label{deltasy}
		\frac{\partial \delta s^y}{\partial t}+\frac{\bf p}{m}\cdot\nabla_{\bf x}\delta s^{y}\approx&\frac{\Gamma_c}{2}\eta({\bf x})\delta J^y\rho({\bf p}),\\
		\label{deltasz}
		\frac{\partial \delta s^z}{\partial t}+\frac{\bf p}{m}\cdot\nabla_{\bf x}\delta s^{z}\approx&0,
	\end{align}
	where we have neglected terms that are second order in the fluctuations. Since Eq.~\eqref{deltasx} and Eq.~\eqref{deltasy} are equivalent, we solve without loss of generality only the equation for $\delta s^x$.
	
	Using the Laplace transformation
	\begin{align}
		\label{laplace}
		L[g](\nu)=\int_{0}^{\infty}\,dt\,e^{-\nu t}g(t),
	\end{align}
	we can find a closed expression for $L[\delta J^x]$ given by
	\begin{align}
		L[\delta J^x]=\frac{\int dt e^{-\nu t}\int d{\bf x}\int d{\bf p}\eta\left({\bf x}+\frac{\bf p}{m}t\right)\delta s^x({\bf x},{\bf p},0)}{D(\nu)},
	\end{align}
	where $\delta J^x=\int d{\bf x}\int d{\bf p}\eta({\bf x})\delta s^x$
	and
	\begin{align}
		\label{Dispersionrelationnonsupp}
		D(\nu) =&1-\frac{\Gamma_c}{2}\int_{0}^{\infty} dt e^{-\nu t}\int d{\bf x}\int d{\bf p}\eta\left({\bf x}+\frac{\bf p}{m}t\right)\eta\rho
	\end{align}
	is the dispersion relation for the NSR phase. The detailed derivation is reported in Appendix~\ref{App:NSR}.
	
	The zeros of the dispersion relation $D(\nu)$ determine the exponents in the time evolution of $\delta J^x$. Assuming that these exponents are negative, the largest exponent (with smallest absolute value) determines the characteristic timescale for a perturbation to relax the spin states again to zero. On the other hand if there exists a zero of the dispersion relation with positive real part, then the NSR phase is unstable. In this case the real part can be seen as the superradiant emission rate.
	
	\subsection{Steady-state superradiant phase (SSR)}
	We will now investigate the mean-field properties of the superradiant phase with a stationary collective dipole. We will refer to the phase as steady-state superradiant (SSR) providing the system reaches a stationary state that fulfills $(J^x,J^y)\neq (0,0)$. Strictly speaking, this is only true in the absence of noise. In the presence of noise, $(J^x,J^y)\neq (0,0)$ is almost always true. In that case steady-state superradiance can be well-characterized by the length of the vector $(J^x,J^y)$ increasing in proportion to the intracavity atom number $N$, i.e., $\|(J^x,J^y)\|\propto N$.
	
	\subsubsection{Analytical solution to the SSR phase}
\begin{figure}[h!]
		\center
		\includegraphics[width=0.6\linewidth]{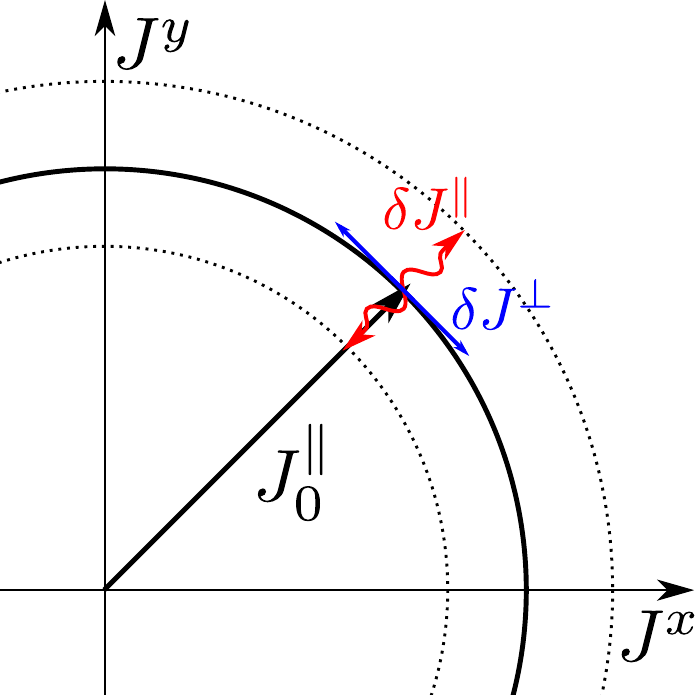}
		\caption{Schematic of the stationary collective dipole in the $J^x$-$J^y$ plane. Its mean length is given by $J^{\parallel}_0$ as defined in Eq.~\eqref{Jparallel}. The dynamics of its length fluctuations, $\delta J^\parallel$, we interpret as a Higgs mode, and the dynamics of its phase fluctuations, $\delta J^\perp$, as a Goldstone mode (see Sec.~\ref{SSR stab}). \label{Fig:Colldipole}}
	\end{figure}

	Our model has an underlying $U(1)$ symmetry as we show in Appendix~\ref{App:U(1)}, therefore this SSR phase can be seen as a symmetry-broken phase~\cite{DeGiorgio:1970}. We can always rotate the system to a frame where the stationary collective dipole $(J^x,J^y)$ points in $x$ direction (see Fig.~\ref{Fig:Colldipole}). We denote the new $x$ axis by $\parallel$ and the perpendicular direction by $\perp$. The resulting equations in the new frame are
	\begin{align}
		\frac{\partial s^\parallel}{\partial t}+\frac{\bf p}{m}\cdot\nabla_{\bf x}s^\parallel=&\frac{\Gamma_c}{2}\eta({\bf x})J^\parallel s^z\label{paras}+\mathcal{S}^\parallel,\\
		\frac{\partial s^{\perp}}{\partial t}+\frac{\bf p}{m}\cdot\nabla_{\bf x}s^{\perp}=&\frac{\Gamma_c}{2}\eta({\bf x})J^{\perp}s^z\label{orths}+\mathcal{S}^\perp,\\
		\label{parsz}
		\frac{\partial s^z}{\partial t}+\frac{\bf p}{m}\cdot\nabla_{\bf x}s^{z}=&-\frac{\Gamma_c}{2}\eta({\bf x})\left(J^\parallel s^\parallel+J^\perp s^\perp\right)+\mathcal{S}^z,
	\end{align}
	with corresponding input noise $W^\parallel$ and $W^\perp$.
	Since the collective dipole points in the $\parallel$ direction, the perpendicular direction $\perp$ is solely noisy with zero mean, implying that $J^{\perp}\approx0$. This leads to the stationary solution for the dipole density $s^{\perp}\approx0$.
	
	Neglecting all noise sources, we can derive the stationary mean-field densities. The mean-field dipole in the perpendicular direction is just $s^{\perp}_0=0$. The mean-field densities $s^{\parallel}_0$ and $s^{z}_0$ are determined by
	\begin{align}
		\label{Eqsparasteady}
		\frac{\bf p}{m}\cdot\nabla_{\bf x}s^{\parallel}_0=&\frac{\Gamma_c}{2}\eta({\bf x})J^{\parallel}_0 s^{z}_0,\\
		\label{Eqzparasteady}
		\frac{\bf p}{m}\cdot\nabla_{\bf x}s^{z}_0=&-\frac{\Gamma_c}{2}\eta({\bf x})J^{\parallel}_0 s^{\parallel}_0,
	\end{align}
	where
	\begin{align}
		\label{Jparallel}
		J^{\parallel}_0=\int d{\bf x}\int d{\bf p}\eta({\bf x})s^{\parallel}_0
	\end{align}
	is the stationary length of the collective dipole.
	Equations~\eqref{Eqsparasteady}--\eqref{Eqzparasteady} can be collected into a single equation
	\begin{align*}
		\frac{\bf p}{m}\cdot\nabla_{\bf x}\left[(s^{\parallel}_0)^2+(s^{z}_0)^2\right]=&0
	\end{align*}
	and therefore solved as
	\begin{align}
		\label{sz0}
		s^{z}_0=\rho({\bf p})\cos[K({\bf x},{\bf p})],\\
		\label{spar0}
		s^{\parallel}_0=\rho({\bf p})\sin[K({\bf x},{\bf p})],
	\end{align}
	where the argument $K({\bf x},{\bf p})$ is determined by
	\begin{align}
		\label{Kdiff}
		\frac{\bf p}{m}\cdot\nabla_{\bf x}K({\bf x},{\bf p})=&\frac{\Gamma_c}{2}\eta({\bf x})J^{\parallel}_0.
	\end{align}
	We will now derive the stability of the SSR phase.
	
	\subsubsection{\label{SSR stab}Stability of the SSR phase}
	Similar to our methods in Sec.~\ref{subsec: non-sr}, we derive the dynamics of small perturbations around the stationary mean-field results by writing the spin densities as $s^{\parallel}=s^{\parallel}_{0}+\delta s^{\parallel}$, $s^{z}=s^{z}_{0}+\delta s^{z}$, and $s^{\perp}=\delta s^{\perp}$. The dynamics of the small fluctuations is governed by the following set of linearized equations
	\begin{align}
		\label{deltaspar}
		\frac{\partial \delta s^{\parallel}}{\partial t}+\frac{\bf p}{m}\cdot\nabla_{\bf x}\delta s^{\parallel}=&\frac{\Gamma_c}{2}\eta({\bf x})\delta J^\parallel s^{z}_0+\frac{\Gamma_c}{2}\eta({\bf x})J^\parallel_0 \delta s^{z},\\
		\label{deltasort}
		\frac{\partial \delta s^{\perp}}{\partial t}+\frac{\bf p}{m}\cdot\nabla_{\bf x} \delta s^{\perp}=&\frac{\Gamma_c}{2}\eta({\bf x}) \delta J_{\perp}s^{z}_0,\\
		\label{delta sz}
		\frac{\partial \delta s^{z}}{\partial t}+\frac{\bf p}{m}\cdot\nabla_{\bf x}\delta s^{z}=&-\frac{\Gamma_c}{2}\eta({\bf x})\left(\delta J^\parallel s^\parallel_0+J^\parallel_0 \delta s^\parallel\right).
	\end{align}
	Notice that using Eq.~\eqref{deltasort} the dynamics of $\delta s^{\perp}$ is completely decoupled from the dynamics of $\delta s^{\parallel}$ and $\delta s^z$. We will rely on this fact to treat the dynamics of these equations separately. Specifically, we interpret the dynamics of $\delta J^{\parallel}$ and $\delta J^{\perp}$ as the Higgs and the Goldstone mode respectively (see Fig.~\ref{Fig:Colldipole}), as we will now elaborate on by examining key aspects of the form of the solutions.
	
	\paragraph{Higgs mode}
	The time evolution of $\delta s^{\parallel}$ together with the coupling to $\delta s^z$ describes the relaxation dynamics of the amplitude of the collective dipole. This can be interpreted as a Higgs mode~\cite{Higgs:1964, Englert:1964}.
	
	Using the Laplace transform defined in Eq.~\eqref{laplace} we can find the following equation
	\begin{align}
		L[\delta J^{\parallel}]=\frac{A^{\parallel}(\nu)}{D_{\parallel}(\nu)},
	\end{align}
	where we have defined $\delta J^{\parallel}=\int d{\bf x}\int d{\bf p}\delta s^{\parallel}$ and the Higgs mode dispersion relation
	\begin{align}
		\label{DispersionHiggs}
		D_{\parallel}(\nu)=&1-\frac{\Gamma_c}{2}\int_{0}^{\infty} dt e^{-\nu t}\int d{\bf x}\int d{\bf p}\eta\left({\bf x}-\frac{\bf p}{m}t\right)\eta s^{z}_0.
	\end{align}
	Details of this derivation and the exact form of $A^{\parallel}(\nu)$ are reported in Appendix~\ref{App:Higgs}.
	We emphasize that in the limit of no superradiance, i.e., $s^{z}_0=\rho$, we obtain the same dispersion relation as we have derived in Eq.~\eqref{Dispersionrelationnonsupp}.
	
	If the SSR phase is stable, we need all the zeros of the dispersion relation $ D_{\parallel}(\nu)$ to have negative real parts. These zeros describe the relaxation dynamics of perturbations in the collective dipole's longitudinal direction.
	
	\paragraph{Goldstone mode}
	The dynamics of $\delta s_{\perp}$ is decoupled from the Higgs mode and describes the evolution of fluctuations perpendicular to it. This is related to the dynamics of the phase of the collective dipole (see Fig.~\ref{Fig:Colldipole}). Because of this observation we refer to this mode as the Goldstone mode~\cite{Goldstone:1961, Goldstone:1962}.
	
	Using the Laplace transform we find
	\begin{align}
		L[\delta J^{\perp}]=\frac{A^{\perp}(\nu)}{D_{\perp}(\nu)},
	\end{align}
	with $\delta J^{\perp}=\int d{\bf x}\int d{\bf p}\delta s^{\perp}$
	and the Goldstone mode dispersion relation
	\begin{align}
		\label{DispersionGoldstone2}
		D_{\perp}(\nu)=& \nu \frac{\int_{0}^{\infty} e^{-\nu t}dt\int d{\bf x}\int d{\bf p}\eta\left({\bf x}+\frac{\bf p}{m}t\right)s^\parallel_0}{J^\parallel_0}.
	\end{align}
	Details of this derivation are shown in Appendix~\ref{App:Goldstone}.
	
	In order for the SSR phase to be stable we require that every zero of Eq.~\eqref{DispersionGoldstone2} cannot have a positive real part. However, we find that the Goldstone dispersion relation always has a zero $\nu=0$ in the SSR phase. This shows that there is no damping of the phase as a consequence of the underlying $U(1)$ symmetry. Every noise will lead to a slight and slow change in $J_{\perp}$. This dynamics is slow compared to the exponents given by the Higgs dispersion relation that determine the relaxation time to the stable length of the collective dipole. However, the slow change in $J_{\perp}$ leads to phase diffusion and this determines the linewidth of the emitted light in the SSR phase~\cite{Lamb:1999} as we will explain in the next section.
	
	\section{\label{sec: analytical}Analytical estimates for the linewidth}
	In the `bad-cavity' regime, where the cavity linewidth exceeds all other frequencies in the system, the coherence is stored in the collective dipole rather than in the cavity field. Therefore the first-order coherence function, $g_1(t)$, for the cavity field is determined by the dipole-dipole correlations
	\begin{align}
		\lim_{t_0\to\infty}\langle \hat{a}^\dag(t+t_0)\hat{a}(t_0)\rangle\propto\lim_{t_0\to\infty}\langle \hat{J}^{+}(t+t_0)\hat{J}^-(t_0)\rangle.
	\end{align}
	In our semiclassical description we exchange the quantum operators for their classical noisy counterparts and correspondingly define the $g_1$ function as
	\begin{align}
		\label{g1}
		g_1(t)=\lim_{t_0\to\infty}\langle J^{*}(t+t_0)J(t_0)\rangle,
	\end{align}
	where we have used $J^{*}=(J^x+iJ^y)/2$ and $J=(J^x-iJ^y)/2$.
	
	\subsection{\label{linewidthnoSR}Linewidth in the NSR phase}
	We first study the behavior of the $g_1$ function in the NSR phase. Here, both dipole components $J^x$ and $J^y$ can be analyzed independently since they are dominated by noise. In this regime we can calculate the $g_1$ function as
	\begin{align}
		g_1(t)\approx\lim_{t_0\to\infty}\frac{\langle J^{x}(t+t_0)J^x(t_0)\rangle+\langle J^{y}(t+t_0)J^y(t_0)\rangle}{4}.
	\end{align}
	Since the noise terms are isotropic, the correlation function for $J^x$ and $J^y$ are the same. Without loss of generality we will focus on the $J^x$ correlation function. For this we define the $g_1^x$ function as
	\begin{align}
		g_{1}^x(t)= \lim_{t_0\to\infty}\langle J^x(t+t_0)J^x(t_0)\rangle.
	\end{align}
	In Appendix~\ref{App:LinewidthNSR} we show that in the long time limit $t\gg\tau$ we find
	\begin{align}
		g_{1}^x(t)\propto e^{\nu_0t},   
	\end{align}
	where $\nu_0$ is the zero with the largest real part of the dispersion relation in Eq.~\eqref{Dispersionrelationnonsupp}. In fact, in the NSR phase, we require that all zeros of Eq.~\eqref{Dispersionrelationnonsupp} are negative. Therefore the $g_1$ function shows an exponential decay on a typical timescale $-1/\mathrm{Re}(\nu_0)$. On the other hand if we approach the transition to the SSR phase we expect that $\mathrm{Re}(\nu_0)$ becomes vanishingly small. This results in a increasing coherence time when approaching the threshold to SSR. 
	
	However, also in the SSR phase, we do not find an actual diverging coherence time. In this phase we have to use a different method to find an estimate for the linewidth as we will now show.
	
	\subsection{Linewidth in the SSR phase}
	The dynamics of $g_1$ and its analysis are very different in the SSR phase. The main difference is that the collective dipole is macroscopic and not dominated by noise. As we have shown in the previous section, we can still decouple two different modes of this dipole, one along the direction of the collective dipole (Higgs mode) and another perpendicular to this direction (Goldstone mode). It is reasonable to write the $g_1$ function in Eq.~\eqref{g1} as
	\begin{align}
		g_1(t)=\lim_{t_0\to\infty}\frac{\langle J^{\parallel}(t+t_0)J^{\parallel}(t_0)e^{i(\varphi(t+t_0)-\varphi(t_0))}\rangle}{4},
	\end{align}
	where we define the collective dipole to be $J(t)=J^{\parallel}(t)e^{-i\varphi(t)}/2$. 
	
	Since the length of the dipole is assumed to be stable, we can always write $J^{\parallel}(t)=J^{\parallel}_0+\delta J^{\parallel}(t)$, where the first term is the stationary length of the collective dipole and $\delta J^{\parallel}(t)$ describes noisy fluctuations around this length (see Fig.~\ref{Fig:Colldipole}). Assuming now that all zeros of the Higgs dispersion relation in Eq.~\eqref{DispersionHiggs} have negative real part, we can conclude that these fluctuations decay rapidly. Therefore, we can simplify the $g_1$ function as
	\begin{align}
		g_1(t)\approx\lim_{t_0\to\infty}\frac{(J_0^{\parallel})^2}{4}\langle e^{i\left[\varphi(t+t_0)-\varphi(t_0)\right]}\rangle.
	\end{align}
	In this picture the dynamics of the $g_1$ function is determined by the dynamics of its phase. The dynamics of the phase can be approximated by
	\begin{align}
		\frac{d\varphi(t)}{dt}\approx\frac{\frac{dJ^{\perp}}{dt}}{J_0^{\parallel}}.
	\end{align}
	With this result it is sufficient to determine the time evolution of $J^{\perp}$. In Appendix~\ref{App:LinewidthSSR} we show that in the limit $t\gg\tau$ we can find the following form for the $g_1$ function
	\begin{align}
		g_1(t)\propto e^{-\frac{\Gamma}{2}t},\label{g1dyn}
	\end{align}
	with a linewidth
	\begin{align}
		\label{Gamma}
		\Gamma=&\frac{4}{\Gamma_cC_{\perp}^2(J_0^{\parallel})^2}+\frac{t_{\mathrm{char}}}{C_{\perp}^2(J_0^{\parallel})^2}.
	\end{align}
	Here, $t_{\mathrm{char}}$ is the characteristic time that has the form
	\begin{align}
		t_{\mathrm{char}}=\int_{-\infty}^{\infty}dt\int d{\bf x}\int d{\bf p}\rho({\bf p})\eta\left({\bf x}+\frac{\bf p}{m}t\right)\eta\left({\bf x}\right)\label{tcharmain}
	\end{align} 
	and the quantity $C_{\perp}$ is defined as
	\begin{align}
		C_{\perp}=\frac{\int_{0}^{\infty} dt\int d{\bf x}\int d{\bf p}\eta\left({\bf x}+\frac{\bf p}{m}t\right)s^\parallel_0}{J^\parallel_0}.
		\label{Cperpmain}\end{align} 
	
	\subsection{Discussion and limitations}
	Here we give an example of the order of magnitude, in particular, regarding the number $N$ of dipoles that effectively interact with the cavity mode. We discuss the behavior of the presented quantities when we increase~$N$. Notice that we  scale $\Gamma_c\propto N^{-1}$ so that $N\Gamma_c$ is considered to be of order $1$. This implies a linear scaling of the maximum output power of the field
	\begin{align}
		\kappa\langle \hat{a}^\dag\hat{a}\rangle\approx \Gamma_c\langle \hat{J}^{+}\hat{J}^{-}\rangle\propto N.
	\end{align}
	
	This choice of scaling allows the dispersion relations given in Eq.~\eqref{Dispersionrelationnonsupp}, Eq.~\eqref{DispersionHiggs}, and Eq.~\eqref{DispersionGoldstone2} to be independent of $N$. Therefore the linewidth in the NSR phase, given by $2\nu_0$, is of order $1$ which is the scaling of the collective linewidth. In the SSR phase, however, we have $J^{\parallel}_0\propto N$ and therefore $(J^{\parallel}_0)^2\propto N^2$ implying a coherent collective dipole. In this regime the linewidth, given in Eq.~\eqref{Gamma}, is of order $\Gamma\propto 1/N$ where we have used that $t_{\mathrm{char}}\propto N$ and $C_{\perp}\propto 1$. This highlights the fact that a macroscopic, coherent collective dipole $\propto N$ is needed for a narrow linewidth that is a factor $N$ smaller than that in the NSR phase.
	
	We remark that the calculation of the $g_1$ function in the NSR phase needs the zero $\nu_0$ of $D(\nu)$ to be sufficiently isolated such that the contribution of exponents with faster decay rate only play a minor role. In general it is possible that $\nu_0$ is complex in that case. Since the dispersion relation is real, there is always a second root $\nu_0^*$ that would need to be included in our calculation. However, this will not affect the decay of the $g_1$ function for very large values of $t$ that is only determined by the real part of $\nu_0$. 
	
	In the SSR phase, our calculation is only valid if every zero of the dispersion relation of the Higgs mode [Eq.~\eqref{DispersionHiggs}] is negative. In this case the decay of the Higgs mode is a factor $N$ faster than the dephasing process determined by $\Gamma$. However, if a zero of Eq.~\eqref{DispersionHiggs} has zero real part, our calculation becomes invalid and predicts an instability of the system. In this situation, the system will be either not superradiant or in a dynamical multi-component superradiant (MCSR) phase, as we will see later in Sec.~\ref{sec: thermal beam}. Such an instability will also occur if there is a solution $\nu_0$ with positive real part to $D_{\perp}(\nu_0)=0$, where $D_{\perp}(\nu_0)$ is the Goldstone dispersion relation [Eq.~\eqref{DispersionGoldstone2}] (see Ref.~\cite{Jaeger:2021}).
	
	\section{\label{sec: thermal beam}A thermal beam traversing the cavity}
	We will now analyze an explicit model in detail. To be specific, we use a cavity mode function that is given by
	\begin{align}
		\label{modefunc}
		\eta({\bf x})=\left[\Theta(x+w)-\Theta(x-w)\right]\cos(kz),
	\end{align}
	where $\Theta(x)$ is the Heaviside step function, $w$ is the cavity mode waist, and $k$ is the wavenumber.  We consider an atomic beam traversing this cavity mode with a constant single velocity $v_x = p_x/m$ and a homogeneous spatial atomic density. The transit time is then $\tau = 2w/v_x$. In the $z$ direction, we assume a Maxwell distribution of velocities. We can thus express $\rho(\bf p)$ as
	\begin{align}
		\label{density}
		\rho = \rho(p_z) = \frac{N}{2w\lambda}\sqrt{\frac{\beta_z}{2m\pi}}e^{-\beta_z\frac{p_z^2}{2m}},
	\end{align}
	where $N$ is the intracavity atom number and $\beta_z$ characterizes the momentum width in the $z$ direction.
	
	\subsection{NSR phase}
\begin{figure}[h!]
		\center
		\includegraphics[width=0.9\linewidth]{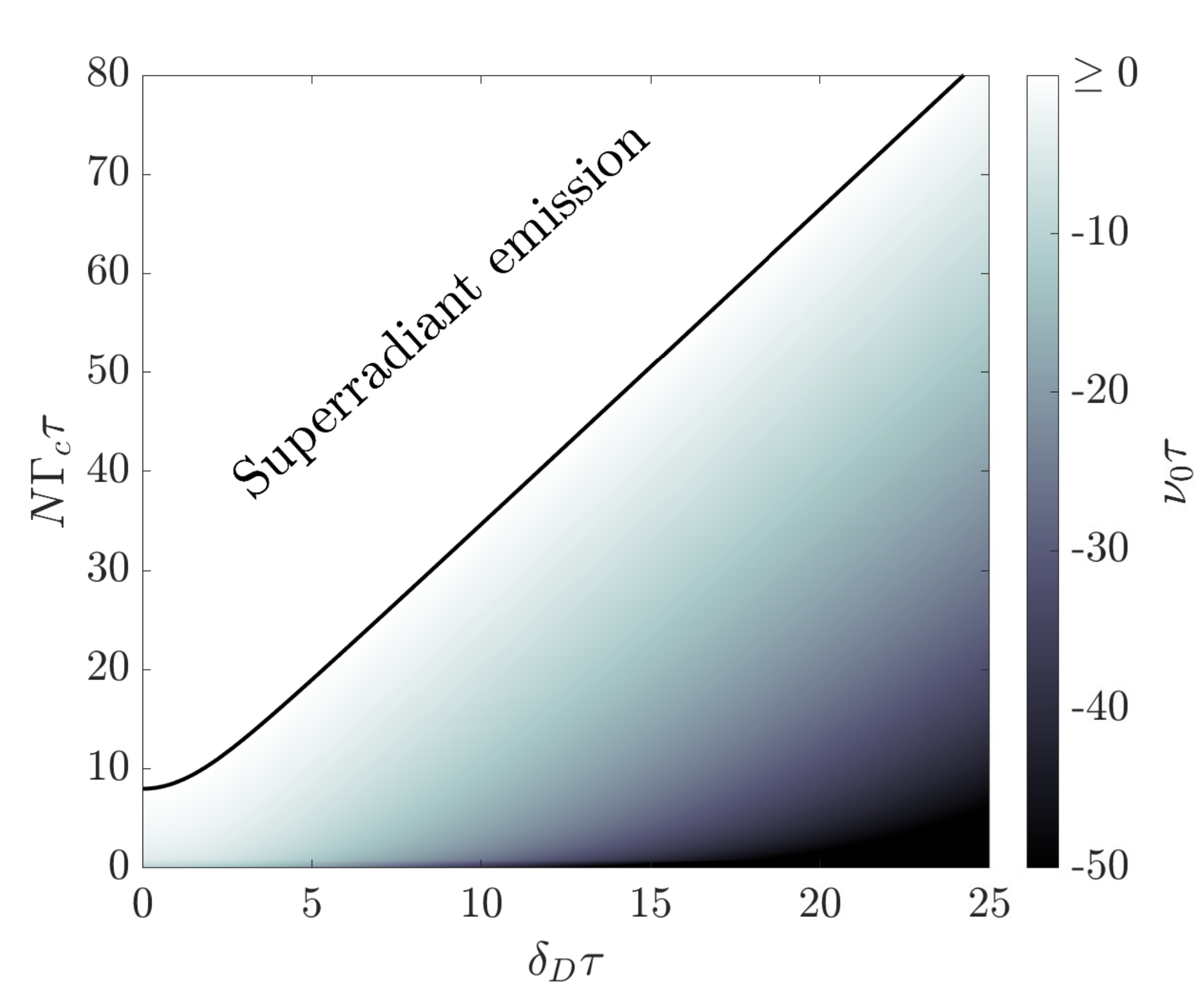}
		\caption{The zero $\nu_0$ of $D(\nu)$ from Eq.~\eqref{Dispersionerf} with the largest real part as a function of the Doppler width $\delta_{D}$ and of the collective linewidth $N\Gamma_c$, all in units of $1/\tau$. In the region where $\nu_0>0$ (shown as white region) the state of the atomic beam is unstable and the beam of excited dipoles will undergo superradiant emission. The solid black line indicates the transition where $\nu_0=0$ [Eq.~\eqref{Dispinhombroad}].\label{Fig:2}}
	\end{figure}

	In the NSR phase all atoms remain in the excited state while they traverse the cavity. The stability of this phase is determined by the dispersion relation in Eq.~\eqref{Dispersionrelationnonsupp}. For the specific case of Eq.~\eqref{modefunc}--\eqref{density}, we can solve the integrals in Eq.~\eqref{Dispersionrelationnonsupp} analytically and obtain
	\begin{align}
		\label{Dispersionerf}
		D(\nu)=&1+\frac{N\Gamma_c\tau}{4}F(\nu),
	\end{align}
	with
	\begin{align*}
		F(\nu)=&\frac{1-e^{-\frac{\delta_{D}^2\tau^2+2\nu\tau}{2}}}{\delta_{D}^2\tau^2}-\sqrt{\frac{\pi}{2\delta_{D}^2\tau^2}}e^{\frac{\nu^2}{2\delta_{D}^2}}\left(1+\frac{\nu\tau}{\delta_{D}^2\tau^2}\right)\nonumber\\
		&\times\left[\mathrm{erf}\left(\frac{\nu+\delta_{D}^2\tau}{\sqrt{2\delta_{D}^2}}\right)-\mathrm{erf}\left(\frac{\nu}{\sqrt{2\delta_{D}^2}}\right)\right].
	\end{align*}
	Here, we have defined the Doppler width as 
	\begin{align}
		\delta_{D} = \frac{k\Delta p_z}{m} = \frac{k}{\sqrt{m\beta_z}},
	\end{align}
	and $\mathrm{erf}(\ldots)$ denotes the error function. The zero $\nu_0$ of Eq.~\eqref{Dispersionerf} with the maximum real part is shown in Fig.~\ref{Fig:2} as a function of $N\Gamma_c\tau$ and $\delta_{D}\tau$.
	For our parameter range solutions are restricted to the domain $\nu_0\in\mathbb{R}$. The shaded area where $\nu_0 < 0$ describes the region where the NSR phase is stable. Here, fluctuations decay with the exponent $\nu_0$. In the white region where $\nu_0 \geq 0$ we expect that fluctuations will be amplified and therefore the atoms will undergo superradiant emission. The condition $\nu_0=0$ describes the phase boundary between the superradiant emission and the NSR phase. This phase boundary can be calculated by solving $D(0)=0$ which results in the equation
	\begin{align}
		\label{Dispinhombroad}
		\frac{N\Gamma_c\tau}{8}=\frac{\delta^2_{D}\tau^2}{\sqrt{2\pi}\delta_{D}\tau\,\mathrm{erf}\left[\frac{\delta_{D}\tau}{\sqrt{2}}\right]+2e^{-\frac{\delta^2_{D}\tau^2}{2}}-2}.
	\end{align}
	We first consider the limit where Doppler broadening is very small, i.e., $\delta_{D}\tau\ll1$. In this case the atoms remain almost in the same position in the standing wave while traversing the cavity. For this choice the right-hand side of Eq.~\eqref{Dispinhombroad} simplifies and we obtain
	\begin{align}
		\label{Dispzerotemp}
		\frac{N\Gamma_c\tau}{8}=1.
	\end{align}
	This shows that even in the absence of Doppler broadening, the collective linewidth $N\Gamma_c$ has to overcome transit-time broadening $1/\tau$, i.e., $ N\Gamma_c > 8/\tau$, so that the atomic beam can induce superradiant emission above threshold.
	
	In the large Doppler broadening limit $\delta_{D}\tau\gg1$, the atoms move many wavelengths during the transit time~$\tau$. In that case, the right-hand side of Eq.~\eqref{Dispinhombroad} can again be simplified, giving
	\begin{align}
		\label{Dispingtemp}
		\frac{N\Gamma_c}{8}=\frac{\delta _{D}}{\sqrt{2\pi}}.
	\end{align}
	This result is a second condition for superradiance; the collective linewidth has to overcome Doppler broadening, i.e., $N\Gamma_c>8\delta_{D}/\sqrt{2\pi}$. Remarkably, this condition is completely independent of $\tau$.
	
	Both conditions $N\Gamma_c>8/\tau$ and $N\Gamma_c>8\delta_{D}/\sqrt{2\pi}$ are visible in Fig.~\ref{Fig:2} in the small ($\delta_{D}\tau\ll1$) and large ($\delta_{D}\tau\gg1$) Doppler broadening limits, respectively. 
	
\begin{figure}[h!]
		\center
		\includegraphics[width=0.65\linewidth]{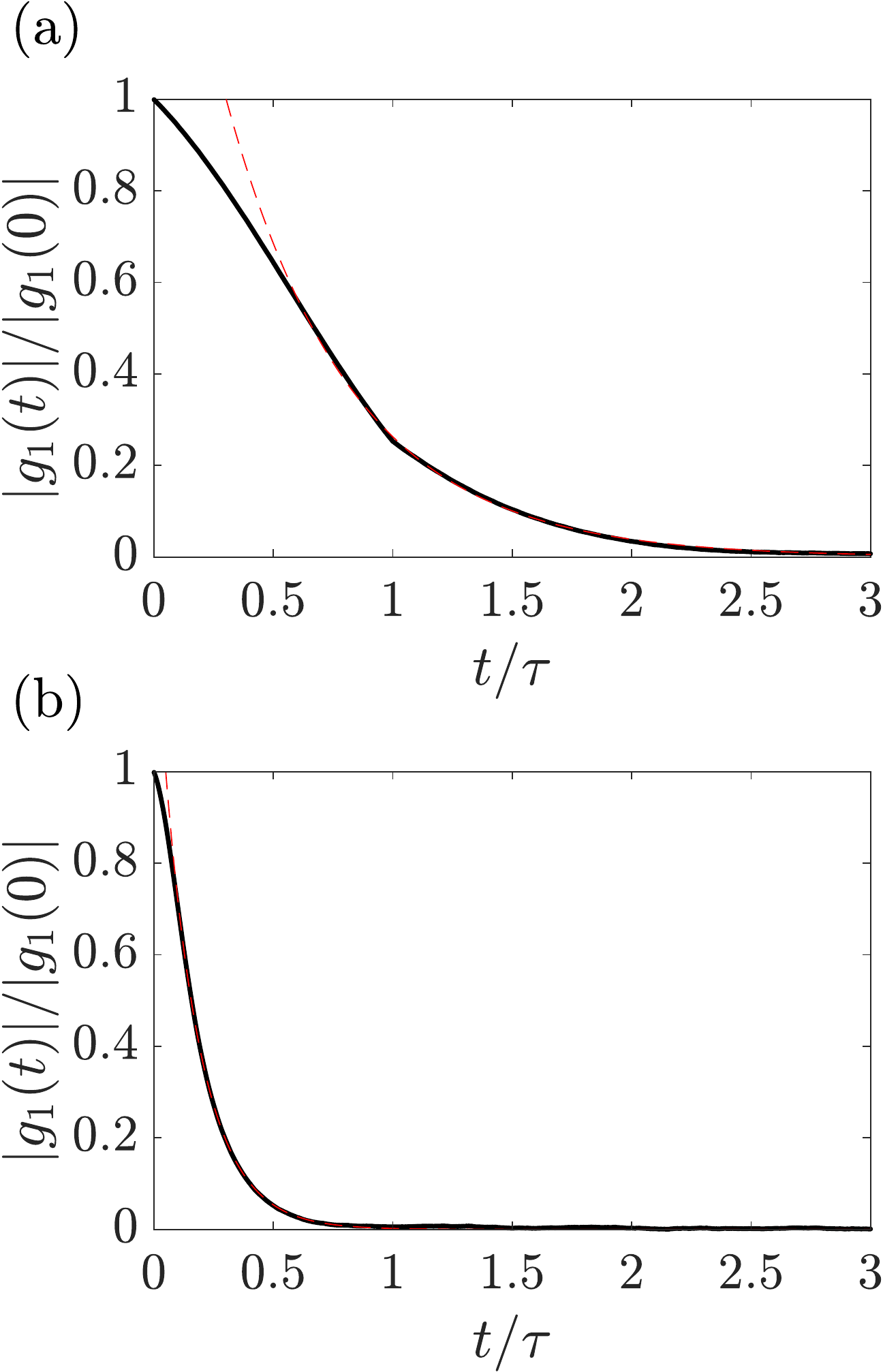}
		\caption{The absolute value of the $g_1$ function [Eq.~\eqref{g1}] normalized by $|g_1(0)|$ as a function of time $t$ in units of $\tau$ for (a) $\delta_D\tau=0.1$, $N\Gamma_c\tau=4$ and (b) $\delta_D\tau=10$, $N\Gamma_c\tau=20$. The $g_1$ function is calculated by numerically integrating Eqs.~\eqref{f}--\eqref{szdensity} using Eqs.~\eqref{modefunc}--\eqref{density} over a total time $t_{\mathrm{sim}}=200\tau$ with ${N=2000}$ atoms, and averaging over $100$ trajectories. For the calculation of $g_1$ we have chosen $t_0=10\tau$. The red dashed line is an exponential fit $\propto \exp(ct)$ of the tail with an exponent $c\tau\approx-1.9$ (a) and $c\tau\approx-6.5$ (b), respectively. The values of $\nu_0$ (see Fig.~\ref{Fig:2}) for the same parameters are $\nu_0\tau=-1.8$ (a), and $\nu_0\tau=-6.2$ (b). \label{Fig:3}}
	\end{figure}

	We will now present results for the $g_1$ function in the NSR phase as defined in Eq.~\eqref{g1} for $t_0\gg \tau$. The analytical estimates of $g_1(t)$ have already been discussed in Sec.~\ref{linewidthnoSR}. Numerically, we find that the $g_1$ function has a non-vanishing imaginary part. However, this imaginary part becomes vanishingly small after averaging over many trajectories. In Fig.~\ref{Fig:3}, we plot the absolute value of the $g_1$ function in (a) for $\delta_{D}\tau=0.1$, $N\Gamma_c\tau=4$  and in (b) for $\delta_{D}\tau=10$, $N\Gamma_c\tau=20$.
	Well inside the NSR phase, these parameters are chosen to represent the case (a) where transit-time broadening dominates Doppler broadening with $\delta_{D}\tau=0.1$, and (b) where Doppler broadening dominates transit-time broadening with $\delta_{D}\tau=10$. For both cases we observe a long-time behavior that is essentially exponential. To show this we have performed a numerical fit to the tail of the $g_1$ function assuming an exponential $\propto \exp(ct)$ and have calculated for (a) $c\tau\approx-1.9$, and for (b) $c\tau\approx-6.5$. Those two values are in very good agreement with the calculated values of $\nu_0$ that are for (a) $\nu_0\tau=-1.8$, and for (b) $\nu_0\tau=-6.2$ (see Sec.~\ref{linewidthnoSR}). However, the short time behavior for both parameter choices is not exponential. In Fig.~\ref{Fig:3}(a) we observe initially an almost linear decay of the $g_1$ function that abruptly ends at the transit time $t=\tau$. The $g_1$ function in Fig.~\ref{Fig:3}(b) shows a Gaussian behavior for short times. The timescale where this Gaussian behavior is visible in much shorter $t<0.1\tau$ in agreement with the timescale expected from the larger Doppler width $t\sim 1/(\delta_{D})=0.1\tau$. The two-stage behavior of the $g_1$ function has the signature of being dominated by single-particle effects for short times and by collective effects, as determined by $\nu_0$, for long times.
	
	In the next subsection we will discuss the superradiant regime.
	
	\subsection{SSR phase}
	For the analysis of the SSR phase we solve the partial differential equation Eq.~\eqref{Kdiff}. The solution is given by
	\begin{align}
		\label{Ksolution}
		K(x-w,z,p_z)=&\frac{\Gamma_cJ^\parallel_0 m}{2kp_z}\left[\sin\left(kz\right)-\sin\left(kz-\frac{kp_z}{mv_x}x\right)\right].
	\end{align}
	This solution has the correct boundary condition $K(-w,z,p_z)=0$ implying that all atomic dipoles are in the excited state when entering the cavity. Substituting Eq.~\eqref{Ksolution} in Eq.~\eqref{spar0} and then calculating $J_0^{\parallel}$ defined in Eq.~\eqref{Jparallel}, we obtain 
	\begin{align}
		J^\parallel_0&=N\int_{-\infty}^{\infty}du\,\frac{e^{-\frac{u^2}{2\delta_{D}^2}}}{\sqrt{2\pi\delta_{D}^2}}\frac{1-\mathcal{J}_0\left[\frac{\Gamma_cJ^\parallel_0\tau}{2}\frac{\sin\left(\frac{u\tau}{2}\right)}{\frac{u\tau}{2}}\right]}{\frac{\Gamma_cJ^\parallel_0\tau}{2}},
	\end{align}
	where $\mathcal{J}_n$ is the Bessel function of the first kind of order $n$. This is a non-linear equation for $J_0^{\parallel}$ that can be simplified by defining the average dipole $j_0^{\parallel}=J_0^{\parallel}/N$ that can be calculated by
	\begin{align}
		\label{jparallel0}
		j^\parallel_0&=\int_{-\infty}^{\infty}du\,\frac{e^{-\frac{u^2}{2\delta_{D}^2}}}{\sqrt{2\pi\delta_{D}^2}}\frac{1-\mathcal{J}_0\left[\frac{N\Gamma_c\tau j^\parallel_0}{2}\frac{\sin\left(\frac{u\tau}{2}\right)}{\frac{u\tau}{2}}\right]}{\frac{N\Gamma_c\tau j^\parallel_0}{2}}.
	\end{align}
	This shows the value of $j^\parallel_0$ is completely determined by the value of $N\Gamma_c\tau$ and $\delta_{D}\tau$. For $j^\parallel_0\neq0$ we obtain a superradiant scaling~\cite{Meiser:2010:1}
	\begin{align}
		\left(J_0^{\parallel}\right)^2=N^2\left(j_0^{\parallel}\right)^2\propto N^2.
	\end{align}
	
\begin{widetext}
		
		\begin{figure}[h!]
			\center
			\includegraphics[width=0.9\linewidth]{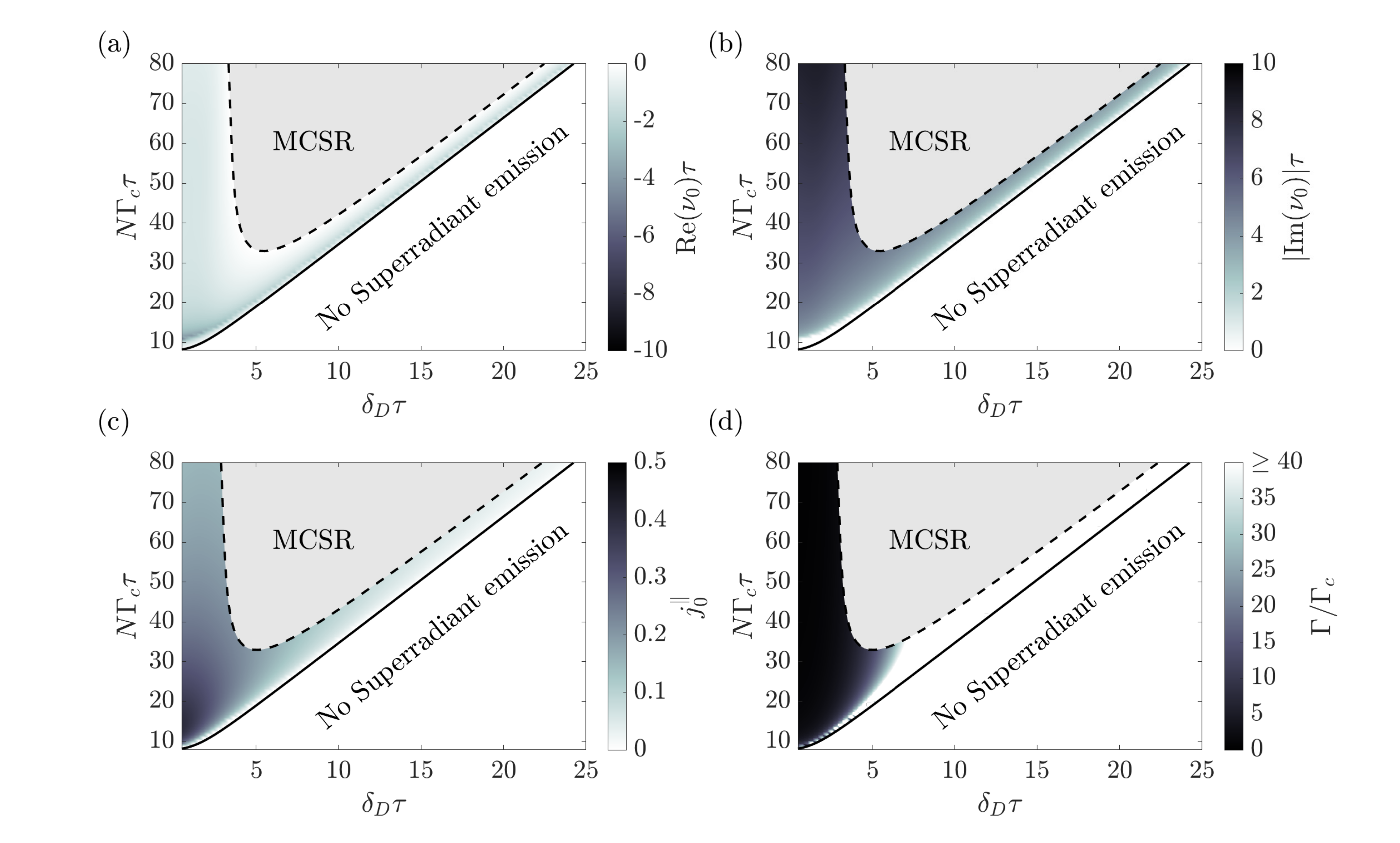}
			\caption{The real part $\mathrm{Re}(\nu_0)$ (a) and the absolute value of the imaginary part $|\mathrm{Im}(\nu_0)|$ (b) in units of $1/\tau$ of the zero $\nu_0$ with the largest real part of the Higgs dispersion relation [Eq.~\eqref{DispersionHiggs}] as a function of the Doppler width $\delta_{D}$ and the collective linewidth $N\Gamma_c$ in units of $1/\tau$. The parameter region where the Higgs mode is unstable, $\mathrm{Re}(\nu_0)>0$, is marked as gray area and bounded by a dashed black line. We call this phase multi-component superradiant (MCSR). The solid black line, given by Eq.~\eqref{Dispinhombroad}, marks the transition from SSR to the NSR phase (see also Fig.~\ref{Fig:2}). Subplots (c) and (d) show the value of the collective dipole $j^{\parallel}_0$ [Eq.~\eqref{jparallel0}] and the linewidth $\Gamma$ [Eq.~\eqref{Gamma}] in units of $\Gamma_c$, respectively. They are shown as a function of the same parameters as subplots (a) and (b) for the parameter regime where the Higgs mode is stable. For all calculations we have used Eq.~\eqref{modefunc} and Eq.~\eqref{density}.\label{Fig:4}}
		\end{figure}
	\end{widetext}

	The stability of this collective dipole is determined by the zero $\nu_0$ with the largest real part of the Higgs and Goldstone mode dispersion relations [Eq.~\eqref{DispersionHiggs} and Eq.~\eqref{DispersionGoldstone2}]. However, for the considered parameter regime we only find an instability in the Higgs mode and not in the Goldstone mode. Because of this, we focus on the Higgs mode dispersion relation in Fig.~\ref{Fig:4}. In order to calculate the zeros of the Higgs dispersion, we substitute Eq.~\eqref{jparallel0} in Eq.~\eqref{Ksolution} to solve for $K({\bf x}, {\bf p})$, and then use Eq.~\eqref{sz0} to calculate the zeros of the dispersion function Eq.~\eqref{DispersionHiggs}. We numerically solve the equation and report the real and imaginary parts of the solution in Fig.~\ref{Fig:4}(a) and (b), respectively.
	We find a parameter regime where $\mathrm{Re}(\nu_0)<0$ and this marks the regime where the SSR phase is stable. However, we observe also an unstable area that is defined by $\mathrm{Re}(\nu_0)>0$. This area is indicated by a gray color in Fig.~\ref{Fig:4} and is bounded by a dashed line that has been determined numerically. In this parameter range we expect neither the NSR nor the SSR phase to be stable. Therefore, we find a dynamical and superradiant behavior of the system that is most clearly visible in the spectrum that has several peaks. Because of this we refer to this phase as multi-component superradiant (MCSR).
	
	In the SSR phase, where $\mathrm{Re}(\nu_0)<0$, we always find a non-vanishing imaginary part $\mathrm{Im}(\nu_0)$ indicating that any fluctuation in the collective dipole length will decay as a damped oscillation. For the whole parameter region of the SSR phase we have also calculated the Goldstone dispersion relation and have not found any additional instabilities.
	
	Figure~\ref{Fig:4}(c) shows the normalized collective dipole $j_0^{\parallel}$ calculated using Eq.~\eqref{jparallel0}. We see that the maximum dipole in the SSR regime is close to $N\Gamma_c\tau=20$ and for $\delta_{D}\tau\ll1$. Using the previous results we can also calculate the linewidth $\Gamma$ using Eq.~\eqref{Gamma}. We expect that this analytical result is valid as long as the collective dipole is stable. The results are apparent in Fig.~\ref{Fig:4}(d). Here, we report a narrow linewidth, $\Gamma<40\Gamma_c$, only for sufficiently small values of $\delta_{D}\tau\lesssim5$.
	
	To analyze and compare our analytical results we have simulated Eqs.~\eqref{f}--\eqref{szdensity} across the different transitions between the SSR, MCSR, and NSR phases.
	
	\subsection{Transition from SSR to NSR}
\begin{figure}[h!]
		\center
		\includegraphics[width=0.95\linewidth]{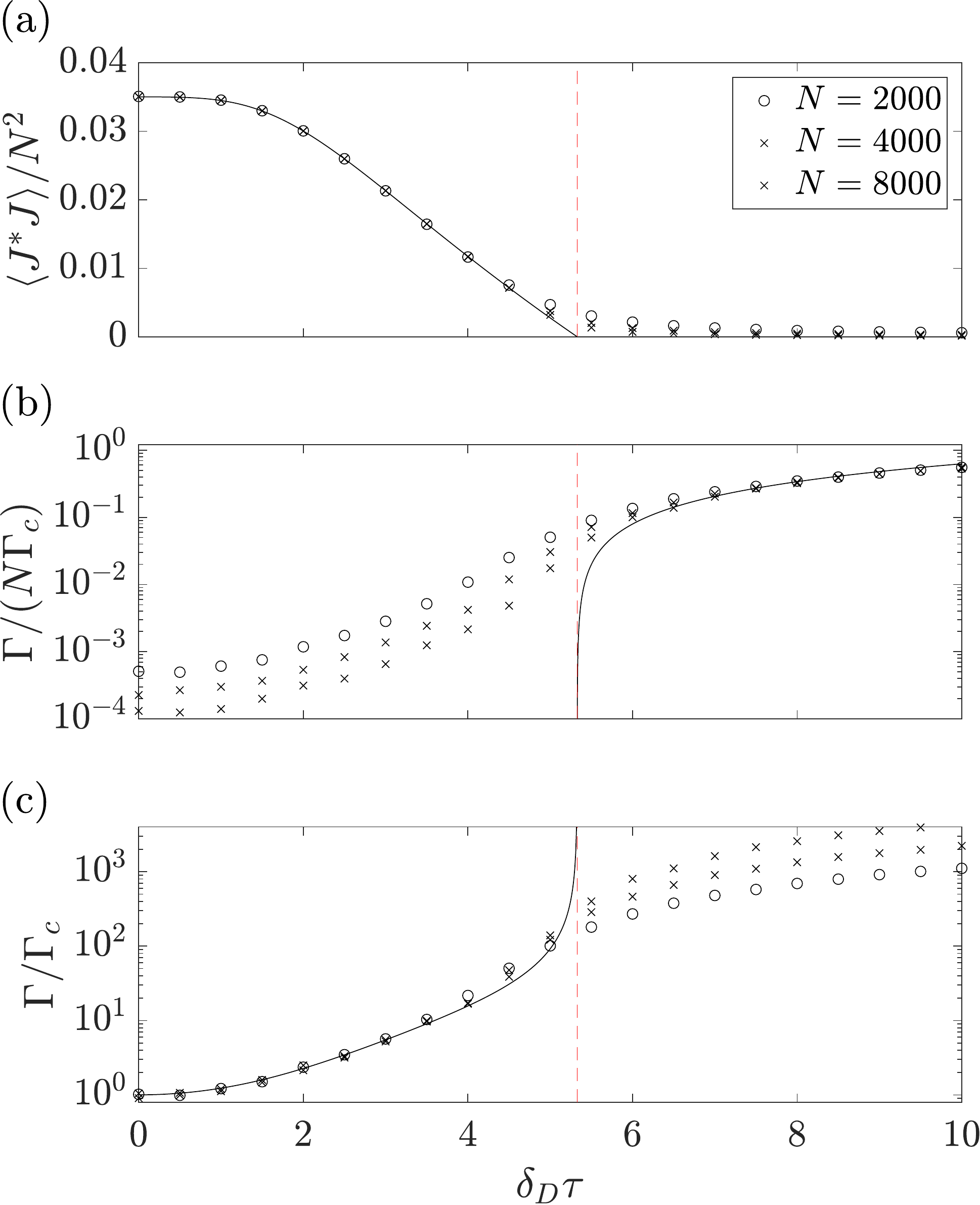}
		\caption{(a) The normalized collective dipole correlation $\langle J^*J\rangle/N^2$, (b) the linewidth $\Gamma$ in units of the collective linewidth $N\Gamma_c$, and (c) the linewidth in units of the single-atom linewidth $\Gamma_c$ as a function of the Doppler width $\delta_D$ in units of $1/\tau$. The different markers correspond to different intracavity atom number $N$ as described in the inset of subplot (a). The linewidth is calculated by fitting the $g_1$ function using $t_0=10\tau$ to an exponential $\propto\exp(-\Gamma t/2)$ over a time interval of length $t_{\mathrm{f}}=20\tau$. The solid line in subplot (a) is the value of $(j_0^{\parallel})^2/4$ calculated from Eq.~\eqref{jparallel0}. The linewidths in (b) visible as solid line are $-2\nu_0$, where $\nu_0$ is the zero with the largest real part of the dispersion relation in Eq.~\eqref{Dispersionerf}. In (c) the solid line gives the solution of Eq.~\eqref{Gamma} calculated using Eq.~\eqref{Ksolution} for given values of $j_0^{\parallel}$. The red dashed vertical lines mark the transition from SSR to the NSR phase. We have chosen $N\Gamma_c\tau=20$ with a simulation time of $t_{\mathrm{sim}}=200\tau$ and a total number of trajectories $200000/N$ for corresponding $N$.  \label{Fig:5}}
	\end{figure}

	We first analyze our simulations for the transition from SSR to the NSR phase for various values of $\delta_D\tau$ and fixed $N\Gamma_c\tau=20$. In Fig.~\ref{Fig:5} we show the results of our numerical integration where different markers indicate different intracavity atom numbers [see inset of Fig.~\ref{Fig:5}(a)].
	
	In Fig.~\ref{Fig:5}(a) we show the collective dipole correlation $\langle J^*J\rangle=\langle (J^x)^2+(J^y)^2\rangle/4$ (proportional to the intensity of the output field) where the red dashed vertical line marks the threshold between the SSR and NSR phases. The analytical prediction is visible as a black solid line and agrees very well with the simulated results. In general we observe that the analytical result is in better agreement for larger intracavity atom number $N$. 
	
	In Fig.~\ref{Fig:5}(b--c) we show the linewidth $\Gamma$ that is extracted by fitting the $g_1$ function in Eq.~\eqref{g1} with $\exp(-\Gamma t/2)$. In subplot (b) the linewidth $\Gamma$ is shown in units of the collective linewidth $N\Gamma_c$ while in subplot (c) we show the linewidth in units of the single-atom linewidth $\Gamma_c$. We observe convergence of the simulation data for different $N$ in the NSR phase in subplot (b). On the other hand we observe convergence of the simulation data in the SSR phase in subplot (c). This finding suggests that the linewidth $\Gamma$ scales with $N\Gamma_c$ in the NSR phase while it scales with $\Gamma_c$ in the SSR phase. 
	
	To further compare our numerical results with analytical predictions we have also calculated the exponent $\nu_0$ that is the zero of the dispersion relation in Eq.~\eqref{Dispersionerf} and plotted it as the black solid line in subplot (b). Numerical and analytical results are in very good agreement in the NSR phase. This description breaks down at the transition where the exponent $\nu_0$ vanishes. After that in the SSR phase we expect that the linewidth of the collectively emitted light is dominated by phase diffusion. In order to show this we have calculated the linewidth in Eq.~\eqref{Gamma} using Eq.~\eqref{Ksolution} and Eq.~\eqref{jparallel0}. This linewidth is plotted as the black line in subplot (c). We find good agreement of the theoretical prediction and the numerical result.
	
	For the derivation of the linewidth in the SSR phase we have assumed a stable length of the collective dipole. This is guaranteed by choosing $N\Gamma_c\tau=20$, where there is no instability in the superradiant regime [see Fig.~\ref{Fig:4}(a)]. In the next subsection we will explicitly study the crossover from the SSR to the MCSR phase, where the Higgs mode becomes unstable.
	
	\subsection{\label{sec: transition}Transition from SSR to MCSR}
\begin{figure}[h!]
		\center
		\includegraphics[width=1\linewidth]{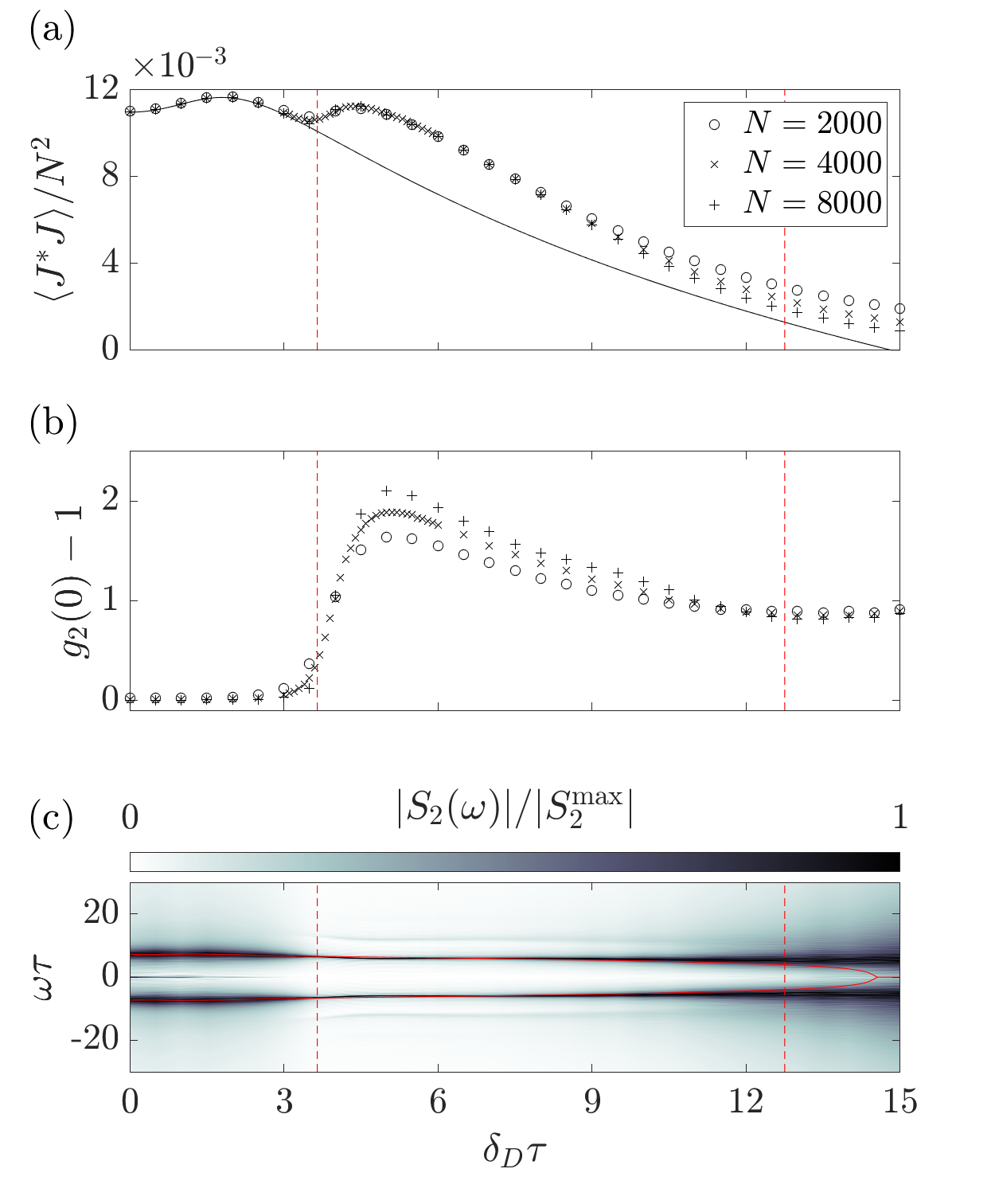}
		\caption{The collective dipole correlation $\langle J^*J\rangle/N^2$ (a) and the value of $g_2(0)-1$ [Eq.~\eqref{g2}] (b) as a function of $\delta_D$ in units of $1/\tau$. The different symbols indicate different intracavity atom numbers $N$ [see inset of subplot (a)]. The solid line in subplot (a) is the value of $(j_0^{\parallel})^2/4$ calculated from Eq.~\eqref{jparallel0}.  Subplot (c) shows the intensity spectrum $|S_2(\omega)|$ defined in Eq.~\eqref{S2omega} as a function of $\omega$ and $\delta_D$ in units of $1/\tau$. The value of $|S_2(\omega)|$ is normalized for every $\delta_D$ by the maximum $|S^{\mathrm{max}}_2| \equiv \mathrm{max}_{\omega}|S_2(\omega)|$ and calculated for $N=4000$. The red vertical dashed lines indicate the threshold from SSR to MCSR and from the MCSR to the SSR phases [see Fig.\ref{Fig:4}(a)]. The red horizontal solid lines in subplot (c) are the values of $\pm\mathrm{Im}(\nu_0)$ corresponding to the zero $\nu_0$ of Eq.~\eqref{DispersionHiggs} with the largest real part. For the calculation of $g_2$ we have used $t_0=10\tau$ and for the calculation of $S_2(\omega)$ and integration time of $t_{\mathrm{f}}=20\tau$. All simulations were performed with $N\Gamma_c\tau=50$ and with a simulation time of $t_{\mathrm{sim}}=200\tau$. For every $N$ we have averaged over $200000/N$ trajectories. \label{Fig:6}}
	\end{figure}

	We choose $N\Gamma_c\tau=50$ and perform simulations for different values of $\delta_D\tau$ across the transition between the SSR and MCSR phases [see Fig.~\ref{Fig:4}(a)]. In Fig.~\ref{Fig:6}(a) we show $\langle J^*J\rangle=\langle (J^x)^2+(J^y)^2\rangle/4$ for different values of $N$ [see inset of Fig.~\ref{Fig:6}(a)]. The red dashed vertical lines mark the thresholds from SSR to the MCSR phase, and from the MCSR to the SSR phase. The first threshold is close to $\delta_D\tau \approx 3$ while the second threshold appears at $\delta_D\tau \approx 12$. For comparison we have also calculated the predicted mean-field value using Eq.~\eqref{jparallel0} that is visible as the black solid line. We find very good agreement in the superradiant phase for small values of $\delta_D\tau$. At the threshold we see an increase of $\langle J^*J\rangle$ in the numerical results that shows a clear deviation from the black line. 
	
	The instability at the transition from SSR to the MCSR phase has been derived from the Higgs dispersion relation that describes the relaxation dynamics of the amplitude of the collective dipole. Therefore we expect to see this instability also in the fluctuations of the collective dipole length. For this we calculate the $g_2$ function which is defined as
	\begin{align}
		\label{g2}
		g_2(t)=\frac{\langle J^{*}(t+t_0)J(t+t_0)J^{*}(t_0)J(t_0)\rangle}{\langle J^{*}J\rangle^2},
	\end{align}
	where $t_0\gg\tau$ is a sufficiently long time. We plot $g_2(0)-1$ in Fig.~\ref{Fig:6}(b) for the same values of $\delta_D\tau$. We find $g_2(0)=1$ well inside the SSR regime ($\delta_D\tau<3$); therefore we expect second-order coherent light. Beyond the transition ($\delta_D\tau\gtrsim3$) we find a sudden increase of $g_2(0)$ highlighting the transition point. This increase cannot be explained by chaotic light because it even exceeds the value of $g_2(0)=2$. Remarkably, the second threshold $\delta_D\tau \approx 12$ is not visible in subplot (b) while we would expect a transition to the SSR phase there with $g_2(0)\approx1$. We understand that this finding is due to finite size effects that are pronounced in this regime because of a small value of $\langle J^*J\rangle/N^2\lesssim 2\times10^{-3}$. This is comparable with finite size effects that we consider to scale like $1/N$.
	
	Because the exponent $\nu_0$ also has an imaginary part [Fig.~\ref{Fig:4}(b)], we also expect an oscillatory behavior in the unstable phase. In order to analyze this we have calculated the intensity spectrum
	\begin{align}
		\label{S2omega}
		S_2(\omega)=\int_{0}^{t_{\mathrm{f}}}dt\,e^{i\omega t}\left[g_2(t)-1\right],
	\end{align}
	where $t_f$ is the integration time. We plot $|S_2(\omega)|$ in Fig.~\ref{Fig:6}(c) as a function of $\omega$ in units of $1/\tau$. The vertical red dashed lines mark the thresholds and the red horizontal solid lines are the values of $\pm\mathrm{Im}(\nu_0)$ visible in Fig.~\ref{Fig:4}(b). We find very good agreement of the values of $\pm\mathrm{Im}(\nu_0)$ with the peaks of $|S_2(\omega)|$ until $\delta_D\tau\lesssim12$.  
	
	The transition between the SSR and the MCSR phase is also visible in Fig.~\ref{Fig:6}(c). The function $|S_2(\omega)|$ shows very broad peaks in the SSR phase suggesting that the amplitude oscillations are strongly damped. This is not true in the MCSR phase where the peaks are narrower suggesting long-lived amplitude oscillations.
	
\begin{figure}[h!]
		\center
		\includegraphics[width=1\linewidth]{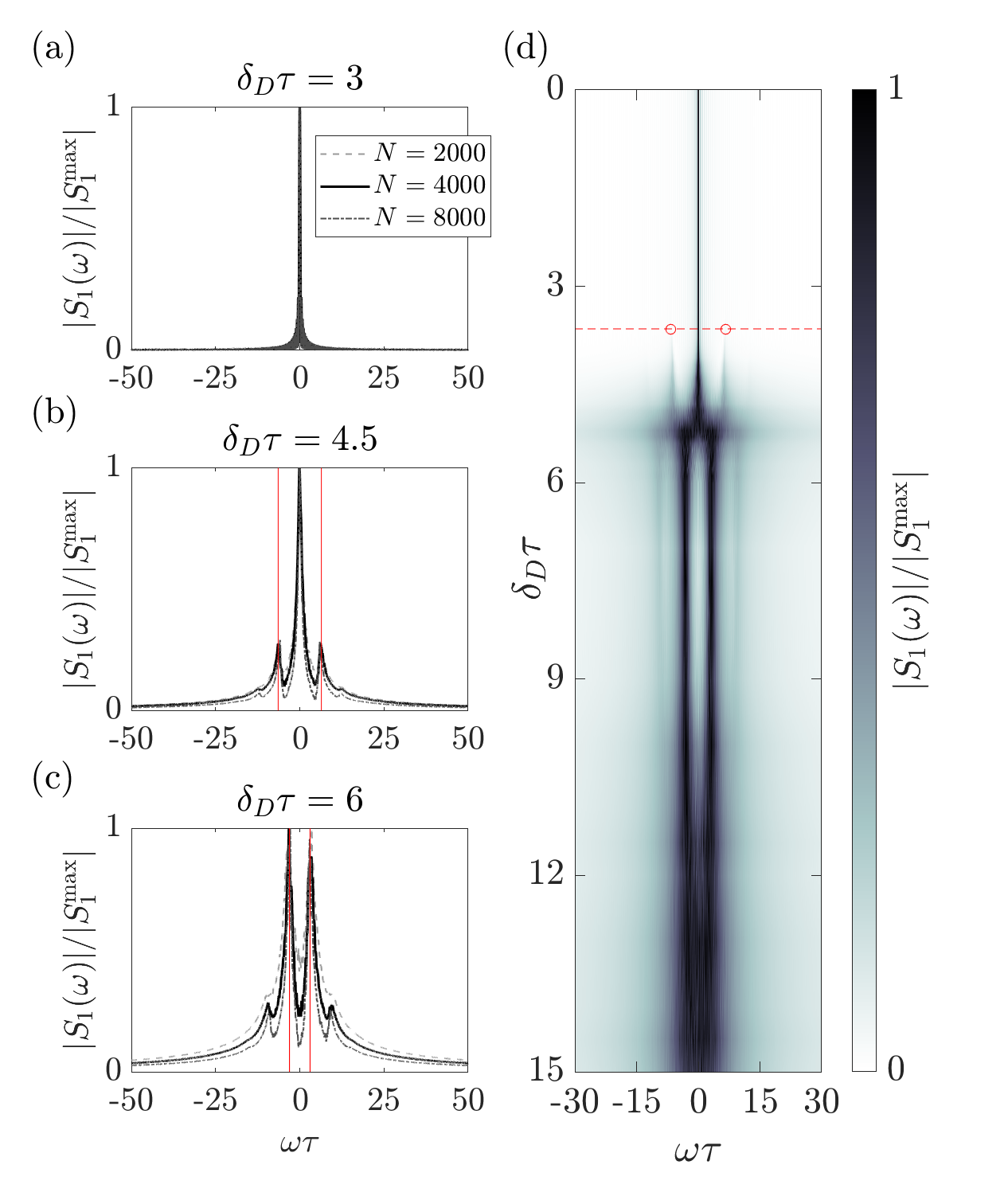}
		\caption{The spectrum $|S_1(\omega)|$ [Eq.~\eqref{S1omega}] plotted for $\delta_D\tau=3$ (a), $\delta_D\tau=4.5$ (b), $\delta_D\tau=6$ (c) as a function of $\omega$ in units of $1/\tau$. The different lines correspond to different intracavity atom numbers $N$ as shown in the inset of subplot (a). The spectrum is normalized for every $\delta_D$ by the maximum $|S^{\mathrm{max}}_1| \equiv \mathrm{max}_{\omega}|S_1(\omega)|$. The red vertical lines in (b) correspond to $\pm\mathrm{Im}(\nu_0)$ where $\nu_0$ is the zero of Eq.~\eqref{DispersionHiggs} with the largest real part. The red vertical lines in (c) correspond to $\pm\mathrm{Im}(\nu_0)/2$. Subplot (d) shows the spectrum $|S_1(\omega)|$ as a function of $\delta_D$ and $\omega$ in units of $\tau$ for $N=4000$. The red dashed horizontal line marks the threshold from the SSR to MCSR regime. The circles on this line are the values of $\pm\mathrm{Im}(\nu_0)$ for the given parameters. All simulations were performed with $N\Gamma_c\tau=50$, with a simulation time of $t_{\mathrm{sim}}=200\tau$ and averaged over $200000/N$ trajectories. The spectra are calculated using $t_0=10\tau$ and $t_{\mathrm{f}}=20\tau$. \label{Fig:7}}
	\end{figure}

	We will study this dynamical feature using the spectrum
	\begin{align}
		S_1(\omega)=\int_{0}^{t_{\mathrm{f}}}dt\,e^{i\omega t}g_1(t),\label{S1omega}
	\end{align}
	which we have calculated for the same parameters (see Fig.~\ref{Fig:7}). Figure~\ref{Fig:7}(d) shows the absolute value of the spectrum $|S_1(\omega)|$ as a function of $\omega$ and $\delta_D$ in units of $1/\tau$. The horizontal dashed red line marks the threshold from SSR to MCSR around $\delta_D\tau \approx 3$. The red circles indicate the value of $\pm\mathrm{Im}(\nu_0)$ at the threshold. In general we find three different appearances in the spectrum:
	\begin{itemize}
		\item[(i)] For sufficiently small values of $\delta_D\tau$ we find one narrow peak at $\omega=0$ indicating coherent and steady-state superradiant emission with the atomic transition frequency. As an example we present a cut of the spectrum in this SSR phase in Fig.~\ref{Fig:7}(a) where we also compare the spectrum for different values of $N$. We remark that in Fig.~\ref{Fig:7}(a) the central peak is Fourier limited because of the finite integration time $t_{\mathrm{f}}$. 
		\item[(ii)] Beyond the transition we find beside the central peak at $\omega=0$ also sidebands. These sidebands appear at the predicted value of $\pm\mathrm{Im}(\nu_0)$. This is also visible in Fig.~\ref{Fig:7}(b) where we have also plotted $\pm\mathrm{Im}(\nu_0)$ as red vertical solid lines for the given parameters. The sidebands become better resolved with increasing $N$. 
		\item[(iii)] Well inside the unstable regime, we find a third behavior where the central peak at $\omega=0$ vanishes and we observe sidepeaks at odd multiples of $\pm\mathrm{Im}(\nu_0)/2$. This is best visible in Fig.~\ref{Fig:7}(c) where we also show $\pm\mathrm{Im}(\nu_0)/2$ as vertical red solid lines corresponding to the given parameters. Here we also find that the peaks become better resolved for increasing $N$. The fact that we find a decreasing width of the sidebands for increasing $N$, as visible in Fig.~\ref{Fig:7}(b--c), suggests that they are due to collective emission.
	\end{itemize}
	
	Remarkably, while the transition from (i)--(ii) is already visible in the length of the collective dipole and the intensity spectrum, the transition (ii)--(iii) is only visible in the coherences that are described by $g_1$. In $g_1$ the peaks occur at $\pm\mathrm{Im}(\nu_0)/2$ while the peaks in $g_2$ are still at $\pm\mathrm{Im}(\nu_0)$. The reason for this is that during an intensity oscillation period $T=2\pi/\mathrm{Im}(\nu_0)$ the collective dipole gains the opposite sign ($J\to-J$). This phase-shift in the collective dipole results in the same intensity ($J^*J\to J^*J$) but doubles the period in $J$ to $2T$. This highlights that the collective dipole is switching between two $\mathbb{Z}_2$ symmetric states in (iii).
	
\begin{figure}[h!]
		\center
		\includegraphics[width=0.9\linewidth]{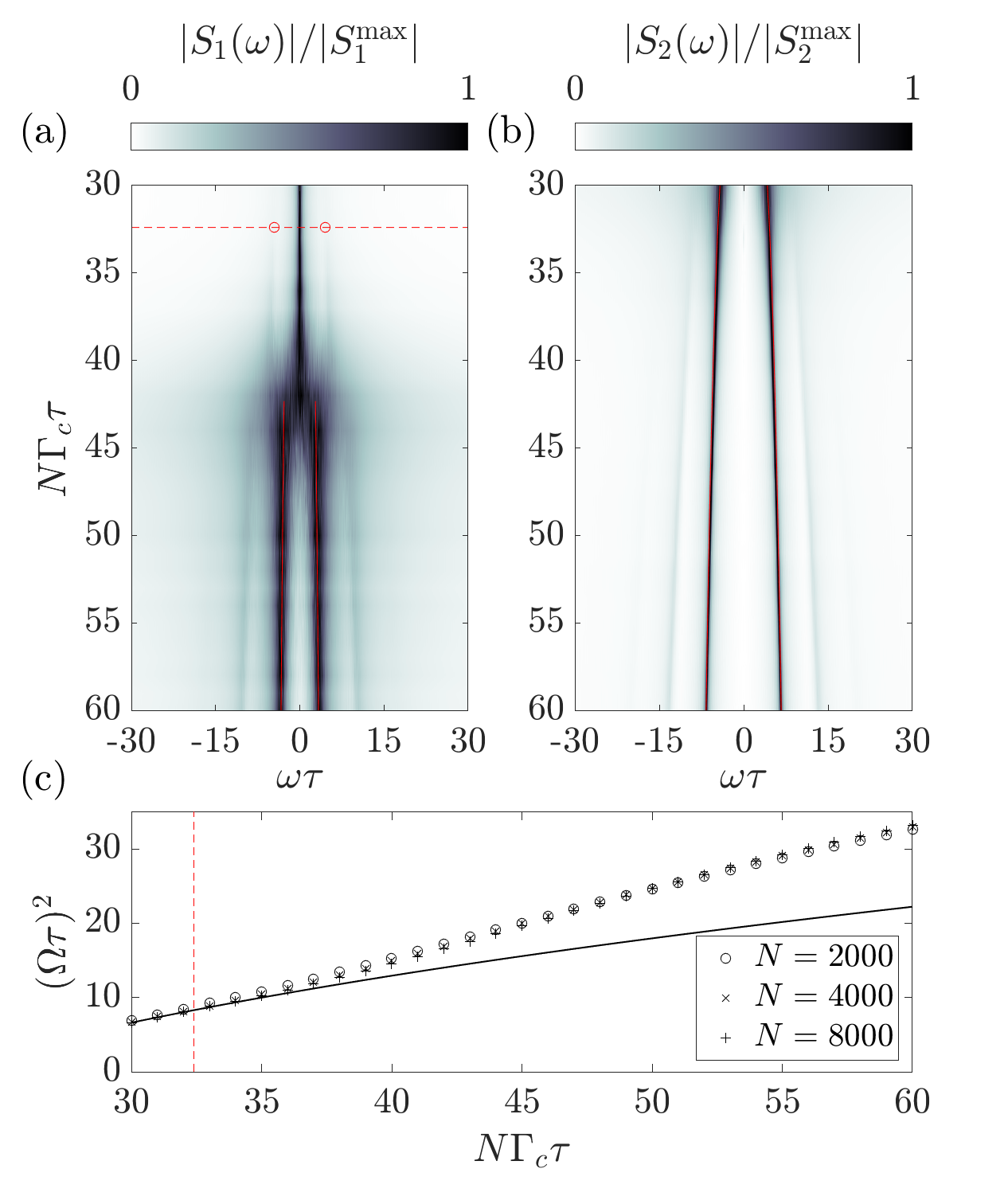}
		\caption{The spectrum $|S_1(\omega)|$ [Eq.~\eqref{S1omega}] (a) and the intensity spectrum $|S_2(\omega)|$ [Eq.~\eqref{S2omega}] (b) as a function of $N\Gamma_c$ and $\omega$ in units of $1/\tau$. Both spectra are normalized for every $\delta_D$ by the maximum $|S^{\mathrm{max}}_n| \equiv \mathrm{max}_{\omega}|S_n(\omega)|$ with $n\in\{1,2\}$. The red dashed horizontal line in (a) marks the threshold between the SSR and the MCSR phase and the circles are the values of $\pm\mathrm{Im}(\nu_0)$. Here, $\nu_0$ is the zero of Eq.~\eqref{DispersionHiggs} with the largest real part. The red solid vertical lines are given by $\pm\mathrm{Im}(\nu_0)/2$. In subplot (b) the red lines show the values of $\pm\mathrm{Im}(\nu_0)$. For all results in subplots (a) and (b) we have used $N=4000$, $t_0=10\tau$, $t_{\mathrm{f}}=20\tau$, and averaged over $50$ trajectories. Subplot (c) shows the squared effective Rabi frequency [Eq.~\eqref{Omega}] in units of $1/\tau^2$ as a function of the collective linewidth $N\Gamma_c$ in units of $1/\tau$. The data are shown for various values of $N$ (see inset). The black solid line shows the result obtained from Eq.~\eqref{jparallel0} and the red vertical dashed line shows the transition from SSR to MCSR. All simulations are performed for $\delta_D\tau=6$.\label{Fig:8}}
	\end{figure}

	To provide further details on this transition we use now a fixed value for the Doppler width $\delta_D\tau=6$ and change the collective linewidth $N\Gamma_c\tau=30$--$60$. For these parameters Fig.~\ref{Fig:4}(a) predicts a transition from SSR to the MCSR phase. The corresponding results for $|S_1(\omega)|$ and $|S_2(\omega)|$ are visible in Fig.~\ref{Fig:8}(a) and Fig.~\ref{Fig:8}(b), respectively.
	The values of $\pm\mathrm{Im}(\nu_0)$ are visible as red lines in Fig.~\ref{Fig:8}(b) and are in good agreement with the sidebands of $|S_2(\omega)|$. We find that the sidebands become narrower when entering the MCSR phase, indicating long-lived intensity oscillations. In the spectrum $|S_1(\omega)|$ in Fig.~\ref{Fig:8}(a) we have marked the theoretically predicted threshold from SSR to MCSR as red dashed horizontal line. The circles on this line show the values of $\pm\mathrm{Im}(\nu_0)$ that agree with the emerging sidebands in $|S_1(\omega)|$. These sidebands become more and more pronounced, emerging from a broad distribution at approximately $N\Gamma_c\tau\approx42$. Beyond this point we find no central peak but a period doubling that we compare to $\pm\mathrm{Im}(\nu_0)/2$ visible as the red lines in Fig.~\ref{Fig:8}(a). We find very good agreement between the sidebands of $|S_1(\omega)|$ and $\pm\mathrm{Im}(\nu_0)/2$ for $N\Gamma_c\tau\gtrsim42$. 
	
	In Fig.~\ref{Fig:8}(c) we show 
	\begin{align}
		\Omega^2 \equiv (N\Gamma_c)^2\langle J^*J\rangle, \label{Omega}
	\end{align}
	which can be seen as the square of an effective Rabi frequency driving the individual dipoles. The quantity is reported in units of $1/\tau^2$ for different intracavity atom numbers [see legend of Fig.~\ref{Fig:8}(c)]. The black solid line is the theoretical prediction obtained from Eq.~\eqref{jparallel0} and is only in good agreement in the SSR phase. The transition between the SSR and the MCSR phases are shown as the vertical red dashed line. We find that the effective Rabi frequency is always larger than the theoretically predicted value.
	
	\subsection{\label{subsec: spontaneous}Spontaneous emission and $T_2$ dephasing}
	We will now discuss the effect of additional noise terms on the observed superradiant phases. In order to do this we study as an example the contribution of free-space spontaneous emission with rate $\gamma_1$ and $T_2$ dephasing with rate $\gamma_2=2/T_2$. We report the dynamical equations that we use to model these processes in Appendix~\ref{App:spontaneous}.
	
\begin{figure}[h!]
		\center
		\includegraphics[width=0.8\linewidth]{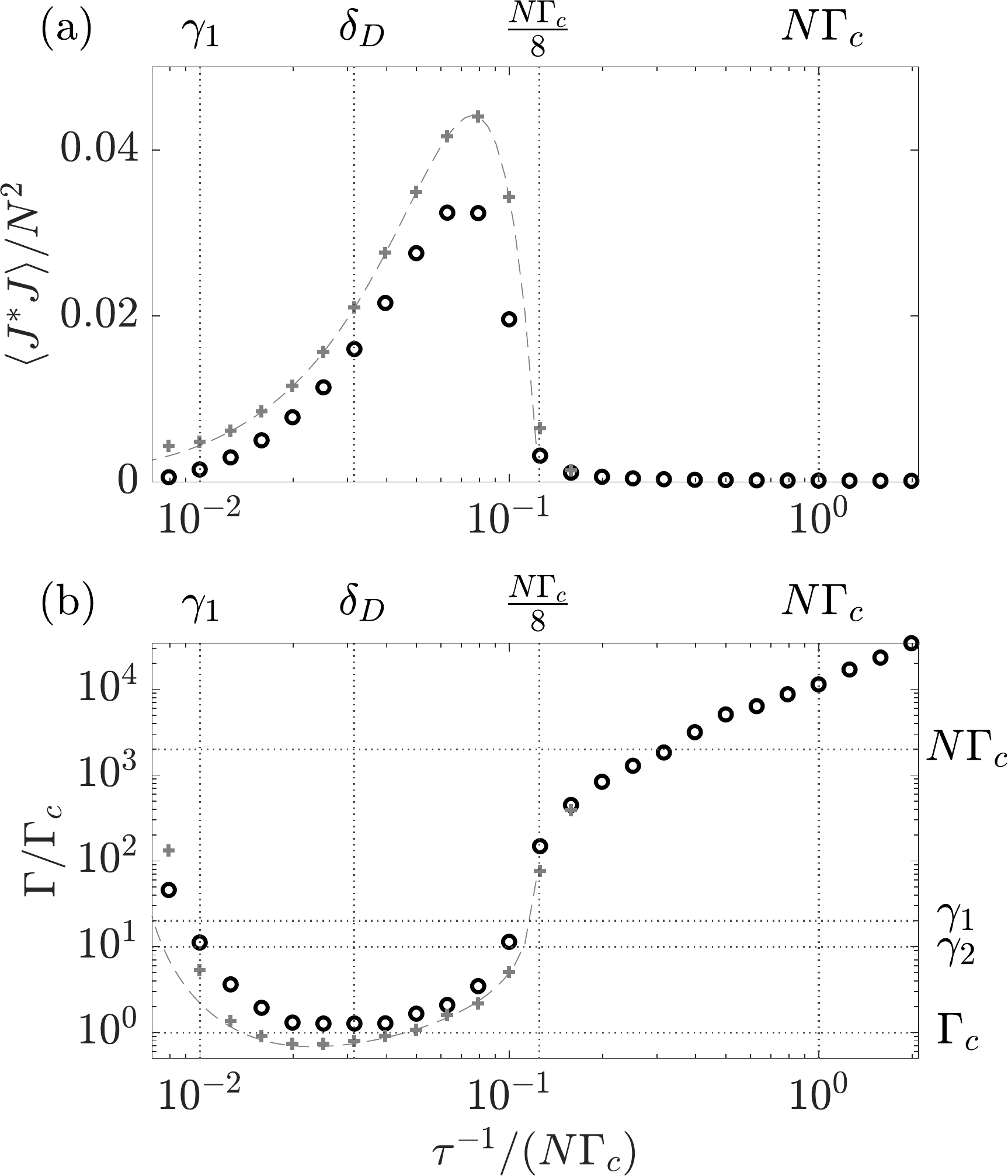}
		\caption{The normalized collective dipole correlation $\langle J^*J\rangle/N^2$ (a) and the linewidth $\Gamma$ in units of the single-atom linewidth $\Gamma_c$ (b) as a function of transit-time broadening $\tau^{-1}$ in units of $N\Gamma_c$. The black circles are simulation results using Eqs.~\eqref{sx_spont}--\eqref{sz_spont}. We have fixed $\delta_D/(N\Gamma_c)=\pi\times10^{-2}$,  $\gamma_1/(N\Gamma_c)=10^{-2}$, $\gamma_2/(N\Gamma_c)=5\times10^{-3}$, and the intracavity atom number $N=2000$. The linewidth is calculated by fitting the $g_1$ function using $t_0=10\tau$ to an exponential $\propto\exp(-\Gamma t/2)$ over a varying $t_{\mathrm{f}}$. All the simulations were performed with $t_{\mathrm{sim}}=100\tau$ and averaged over 100 trajectories. The grey plus symbols are simulation results using the same parameters except for $\gamma_1=0=\gamma_2$. The grey dashed lines are analytical solutions, giving in (a) the value of $(j_0^\parallel)^2/4$ using Eq.~\eqref{jparallel0}, and in (b) the linewidth Eq.~\eqref{Gamma} calculated using Eq.~\eqref{Ksolution} with corresponding values of $j_0^\parallel$, respectively.\label{Fig:9}}
	\end{figure}

	We first investigate how these noise sources affect the SSR phase and in particular the intensity and the linewidth of the produced light. In particular we focus on the regime where the collective linewidth is much larger than the Doppler width $\delta_D/(N\Gamma_c)=\pi\times10^{-2}$, the spontaneous emission rate $\gamma_1/(N\Gamma_c)=10^{-2}$, and the dephasing $\gamma_2/(N\Gamma_c)=5\times10^{-3}$. We fix the intracavity atom number $N=2000$ and vary the ratio between $\tau^{-1}$ and $N\Gamma_c$. In Fig.~\ref{Fig:9}(a) we show the results of $\langle J^*J\rangle/N^2$ for these parameters as black circles. For comparison we have performed simulations with $\gamma_1=0=\gamma_2$ visible as grey pluses and also plotted the analytical result corresponding to the solution of Eq.~\eqref{jparallel0} as grey dashed line. While we find almost perfect agreement between the analytical result and the simulation with $\gamma_1=0=\gamma_2$, the numerical results including spontaneous emission is always smaller. This can be expected because spontaneous emission and dephasing will both result in a decrease of coherence in the atomic dipoles and therefore result in a reduced light intensity. In addition, free-space spontaneous emission also leads to a loss of excitations into electromagnetic modes external to the cavity mode. Nevertheless, we find very good agreement for the threshold of superradiance that for the considered parameter regime is close to $\tau^{-1}/(N\Gamma_c)=1/8$. We also find a similar functional behavior of $\langle J^*J\rangle/N^2$ for the simulations with and without spontaneous emission and dephasing.
	
	Figure~\ref{Fig:9}(b) shows the linewidth $\Gamma$ calculated by fitting the $g_1$ function given by Eq.~\eqref{g1} with $\exp(-\Gamma t/2)$ obtained from simulations including (black circles) and without spontaneous emission and dephasing (grey pluses). We also compare our results to the analytical estimate from Eq.~\eqref{Gamma} visible as grey dashed line. We find very good agreement between the simulations without spontaneous emission and dephasing and the analytical result as long as $\tau^{-1}/(N\Gamma_c)>10^{-2}$. Below this point we find a rather small coherent collective dipole component and cannot expect that the phase diffusion argument that has been used to derive the analytical result will still be valid. The simulations including spontaneous emission show a very similar functional dependence but are almost always slightly above the simulation results without spontaneous emission. Still, we find a minimum linewidth of the order of $\Gamma_c$ that is order of magnitudes smaller than $\gamma_1$ and $\gamma_2$. This highlights the fact that the linewidth of the generated light is typically not limited by any single-particle dephasing mechanism.  
	
\begin{figure}[h!]
		\center
		\includegraphics[width=0.8\linewidth]{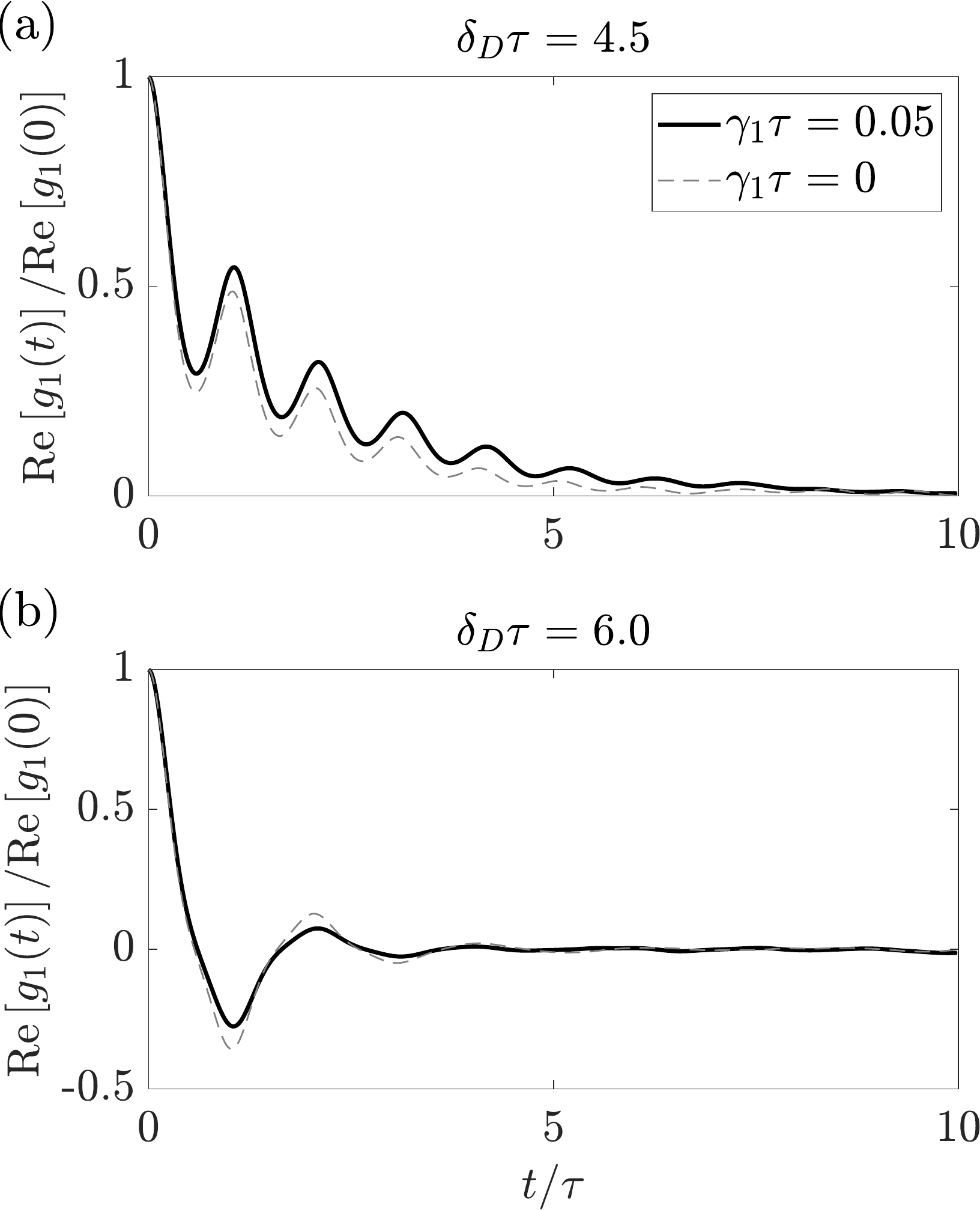}
		\caption{Simulation results of the real part of $g_1(t)$ normalized by $\mathrm{Re}\left[g_1(0)\right]$ for $\delta_D\tau = 4.5$ (a) and $\delta_D\tau = 6.0$ (b). For the black solid lines we have used $N\Gamma_c\tau = 50$, $\gamma_1 = 0.05\tau^{-1}$ with $N=4000$ and $t_{\mathrm{sim}} = 200$. The $g_1$ function is calculated using $t_0=10\tau$ and averaged over 50 trajectories. For the grey dashed lines we have used the same parameters except for $\gamma_1=0$. These dashed lines are the real parts of the $g_1$ functions that are used to calculate the spectra shown in Fig.~\ref{Fig:7} (b) and (c).  \label{Fig:10}}
	\end{figure}

	We will now study the stability of the MCSR phase. For this we choose the same parameters where we have observed the two different emission regimes in Fig.~\ref{Fig:7}(b--c), i.e., $N\Gamma_c\tau=50$, $\delta_D\tau = 4.5$ and $\delta_D\tau =6.0$, respectively. We now add a small spontaneous emission rate $\gamma_1\tau=0.05$ to our previous simulations. 
	We plot the real part of the $g_1$ function $\mathrm{Re}(g_1)$ in Fig.~\ref{Fig:10} for $\delta_D\tau=4.5$ (a) and $\delta_D\tau=6.0$ (b). The simulations without spontaneous emission are visible as grey dashed lines and the simulations with spontaneous emission as black solid lines.
	In Fig.~\ref{Fig:10}(a) we find a positive $\mathrm{Re}(g_1)$ with oscillations for both simulation types that are in good agreement. As a consequence, we also find a similar spectrum as shown in Fig.~\ref{Fig:7}(b). Remarkably, our simulation results suggest that the oscillations have a slightly longer lifetime for non-vanishing $\gamma_1$. 
	
	Figure~\ref{Fig:10}(b) shows very good agreement between the two simulations with and without spontaneous emission. We find $\mathrm{Re}(g_1)$ oscillating around zero, therefore giving rise to a similar spectrum as in Fig.~\ref{Fig:7}(c). Our findings show that the change of the sign in $\mathrm{Re}(g_1)$ that occurs with half the frequency of the intensity oscillations is robust against small additional noise sources.

	\section{\label{sec: conclusion}Conclusions}
	In this paper we have studied the onset and stability of collective emission of an atomic beam that traverses an optical cavity. We have developed a semiclassical theoretical framework to study the dynamics of the atomic dipoles in the presence of Doppler broadening. We have analyzed this model using a mean-field description and determined the stability of the non-superradiant (NSR) and steady-state superradiant (SSR) phases. These results were used to analyze the stationary light emission of the corresponding phases and predict a linewidth of the emitted light. After that we investigated a model using numerical simulations and presented analytical techniques that provide supporting analysis. We explored a SSR phase and a dynamical superradiant phase with a multi-component superradiant (MCSR) light output. With our derived theory we were able to quantitatively predict the threshold of the MCSR phase and the occurrence of sidebands in the spectra. In addition, we found that these results are robust against free-space spontaneous emission and $T_2$ dephasing processes if they are small compared to transit-time broadening and Doppler broadening.
	
	We highlight that the MCSR phase is observed in presence of relatively large Doppler broadening. This is potentially easier to realize in actual experimental setups working with thermal atomic beams. Nevertheless, for the observation of the MCSR phase one still requires a collective linewidth that overcomes all broadening mechanism including Doppler broadening. 
	
	We have focused on the interplay between collective emission and thermal broadening in the parameter regime where thermal effects dominate dephasing processes such as free-space spontaneous emission. However, we expect that these effects become important for cold or even ultracold atomic beams when the Doppler broadening becomes comparable to the linewidth of the atomic dipoles. In this parameter regime one could potentially study subradiance in the regime where the transit time becomes comparable to the atomic lifetime~\cite{Temnov:2005, Shankar:2021}. Additionally, one could explore the regime where the collective linewidth becomes comparable to the recoil frequency~\cite{Bonifacio:1994a, Bonifacio:1994b, Bonifacio:2005, Slama:2007, Jaeger:2019, Jaeger:2020} and the semiclassical theory used in this work becomes invalid. Such parameter regimes could be achievable regarding the recent progress on producing high phase-space density atomic beams~\cite{Chen:2019}.
	
	\section*{Acknowledgments}
	This research is supported by the Research Centres of Excellence program supported by the National Research Foundation (NRF) Singapore; the Ministry of Education, Singapore; the NSF AMO Grant No. 1806827; NSF PFC Grant No. 1734006; and the DARPA and ARO Grant No. W911NF-16-1-0576.
	
	S. B. J{\"a}ger and H. Liu contributed equally to this work.

	\appendix
	\section{\label{App:OptoForces} Neglecting optomechanical forces}
	Our theoretical description is valid if optomechanical forces can be neglected. In this section we discuss the validity of this approximation.
	
	Optomechanical forces are described in Eq.~\eqref{hatp}. In order to justify the approximation of a ballistic motion, we estimate the mean force ${\bf F}_{\mathrm{mean}} \sim \hbar N\Gamma_c\nabla_{\bf x}\eta({\bf x})$ from Eq.~\eqref{hatp} and the mean  momentum change ${\bf F}_{\mathrm{mean}}\tau$, where $\tau \equiv 2w/\langle v_x \rangle$ is the transit time. Here $w$ is the cavity waist and $\langle v_x \rangle = \langle p_x \rangle / m$ is the mean atomic velocity in $x$ direction. The mean momentum change has to be compared with the momentum widths of the initial atomic momentum distribution in the corresponding Cartesian coordinates. Along the $z$ axis, assuming a standing wave potential with wavenumber $k = 2\pi/\lambda$, optomechanical forces are negligible if $\hbar k N\Gamma_c\tau \ll \Delta p_z$, where $\Delta p_z$ is the momentum width in $z$ direction. For $N\Gamma_c\tau \gtrsim 1$ this requires a momentum width that is much larger than the a single photon recoil $\hbar k$. Vertical to the cavity axis, the mean force can be roughly approximated by $\hbar w^{-1}N\Gamma_c$. The condition reads then $\hbar w^{-1}N\Gamma_c\tau\ll\Delta p_y$ and $\hbar w^{-1}N\Gamma_c\tau\ll\langle p_x\rangle$. Therefore, we conclude that optomechanical forces are negligible as long as the temperature of the incoming atoms is sufficiently high.

	\section{\label{App:NSR}Stability of the NSR phase}
	In this section we present the derivation of the dispersion relation for the NSR phase given in Eq.~\eqref{Dispersionrelationnonsupp}.
	
	Applying the Laplace transform [Eq.~\eqref{laplace}] on Eq.~\eqref{deltasx}, we obtain
	\begin{align}
		\label{L_deltasx}
		\left[\nu-\mathcal{L}_0\right]L[\delta s^x]=\delta s^x({\bf x},{\bf p},0)+\frac{\Gamma_c}{2}\eta({\bf x})\rho({\bf p})L[\delta J^x],
	\end{align}
	where we have used the definition
	\begin{align}
		\label{L0}
		\mathcal{L}_0g({\bf x}) = -\frac{\bf p}{m}\cdot\nabla_{\bf x}g({\bf x}).
	\end{align}
	Next we multiply Eq.~\eqref{L_deltasx} first by the inverse of $\left[\nu-\mathcal{L}_0\right]$ and then by $\eta({\bf x})$. After an integration over space and momentum, we obtain a linear equation for $L[\delta J^x]$. This linear equation can be solved to find the result
	\begin{align}
		L[\delta J^x]=\frac{\int d{\bf x}\int d{\bf p}\eta({\bf x})\left[\nu-\mathcal{L}_0\right]^{-1}\delta s^x({\bf x},{\bf p},0)}{1-\frac{\Gamma_c}{2}\int d{\bf x}\int d{\bf p}\eta({\bf x})\left[\nu-\mathcal{L}_0\right]^{-1}\eta({\bf x})\rho({\bf p})}.
	\end{align}
	The denominator is the dispersion function $D(\nu)$ and takes the form
	\begin{align*}
		D(\nu) = &1-\frac{\Gamma_c}{2}\int_{0}^{\infty}dt e^{-\nu t}\int d{\bf x}\int d{\bf p}\eta({\bf x})e^{\mathcal{L}_0t}\eta({\bf x})\rho({\bf p}).
	\end{align*}
	Now using the action of the propagator
	\begin{align}
		e^{\mathcal{L}_0t}f({\bf x})=f\left({\bf x}-\frac{\bf p}{m}t\right),\label{propagator}
	\end{align}
	and after performing a change of variables ${\bf x}\mapsto {\bf x}+{\bf p}t/m$ we obtain the form given by Eq.~\eqref{Dispersionrelationnonsupp}.
	
	\section{\label{App:U(1)}$U(1)$ symmetry of the model}
	In this section we show that Eqs.~\eqref{sxdensity}--\eqref{szdensity} as well as their mean-field versions Eqs.~\eqref{meanfield sx}--\eqref{meanfield sz} have a $U(1)$ symmetry. This symmetry is given by a rotation with an arbitrary $\varphi\in\mathbb{R}$,
	\begin{align}
		\label{U1transformation}
		\begin{pmatrix}
			s^x\\
			s^y
		\end{pmatrix}
		=
		\begin{pmatrix}
			\cos\varphi&-\sin\varphi\\
			\sin\varphi&\cos\varphi\\
		\end{pmatrix}
		\begin{pmatrix}
			\tilde{s}^x\\
			\tilde{s}^y
		\end{pmatrix}
	\end{align}
	that transforms Eqs.~\eqref{sxdensity}--\eqref{sydensity} to
	\begin{align}
		\label{sx'}
		\frac{\partial \tilde{s}^x}{\partial t}+\frac{\bf p}{m}\cdot\nabla_{\bf x}\tilde{s}^x=&\frac{\Gamma_c}{2}\eta({\bf x})\tilde{J}^xs^z+\tilde{\mathcal{S}}^x\\
		\label{sy'}
		\frac{\partial \tilde{s}^y}{\partial t}+\frac{\bf p}{m}\cdot\nabla_{\bf x}\tilde{s}^y=&\frac{\Gamma_c}{2}\eta({\bf x})\tilde{J}^ys^z+\tilde{\mathcal{S}}^y
	\end{align}
	with corresponding noisy initial conditions $\tilde{W}^x$ and $\tilde{W}^y$.  Here, all objects labeled by $\tilde{(\,.\,)}$ are transformed according to the linear operation in Eq.~\eqref{U1transformation}.
	
	\section{\label{App:Higgs}Stability of the SSR phase: Higgs mode}
	In this section we provide details for the derivation of the Higgs mode dispersion relation given by Eq.~\eqref{DispersionHiggs}.
	
	In order to derive this dispersion relation, we first define $\delta s^{+}=\delta s^{\parallel}+i\delta s^{z}$ and $\delta s^{-}=\delta s^{\parallel}-i\delta s^{z}$. We can then use Eq.~\eqref{deltaspar} and Eq.~\eqref{delta sz} to derive two decoupled equations
	\begin{align*}
		\frac{\partial \delta s^{+}}{\partial t}+\frac{\bf p}{m}\cdot\nabla_{\bf x}\delta s^{+}=&-i\frac{\Gamma_c}{2}\eta J^\parallel_0 \delta s^{+}+\frac{\Gamma_c}{2}\rho({\bf p})\eta\delta J^\parallel e^{-iK},\\
		\frac{\partial \delta s^{-}}{\partial t}+\frac{\bf p}{m}\cdot\nabla_{\bf x}\delta s^{-}=&i\frac{\Gamma_c}{2}\eta J^\parallel_0 \delta s^{-}+\frac{\Gamma_c}{2}\rho({\bf p})\eta \delta J^\parallel e^{iK},
	\end{align*}
	where we have used the notations $K=K({\bf x},{\bf p})$, $\rho=\rho({\bf p})$, and $\eta=\eta({\bf x})$.
	These equations can be solved using the Laplace transform given by Eq.~\eqref{laplace} and we find
	\begin{align}
		\label{Ls+}
		\left[\nu-\mathcal{L}_1\right]L[\delta s^{+}]=&\delta s^{+}({\bf x},{\bf p},0)+\frac{\Gamma_c}{2}\rho L[\delta J^{\parallel}]\eta e^{-iK},\\
		\label{Ls-}
		\left[\nu-\mathcal{L}_2\right]L[\delta s^{-}]=&\delta s^{-}({\bf x},{\bf p},0)+\frac{\Gamma_c}{2}\rho L[\delta J^{\parallel}]\eta e^{iK},
	\end{align}
	where
	\begin{align}
		\mathcal{L}_1g({\bf x})=-\frac{\bf p}{m}\cdot\nabla_{\bf x}g({\bf x})-i\frac{\Gamma_c}{2}\eta({\bf x})J^\parallel_0 g({\bf x}),\\
		\mathcal{L}_2g({\bf x})=-\frac{\bf p}{m}\cdot\nabla_{\bf x}g({\bf x})+i\frac{\Gamma_c}{2}\eta({\bf x})J^\parallel_0 g({\bf x}).
	\end{align}
	We can now solve Eqs.~\eqref{Ls+}--\eqref{Ls-} formally for $L[\delta s^{+}]$ and $L[\delta s^{-}]$. Using $L[\delta s^{\parallel}]=(L[\delta s^{+}]+L[\delta s^{-}])/2$, multiplying this expression by $\eta({\bf x})$, and integrating over the whole phase space, we end up with an expression for $L[\delta J_{\parallel}]$. Solving that equation for $L[\delta J_{\parallel}]$ leads to the final expression given by
	\begin{align}
		\label{Dispersionrelationsup}
		L[\delta J^{\parallel}]=\frac{A^{\parallel}(\nu)}{D_{\parallel}(\nu)},
	\end{align}
	with
	\begin{align}
		A^{\parallel}(\nu)=& \frac{1}{2}\int d{\bf x}\int d{\bf p}\eta({\bf x})\left[\nu-\mathcal{L}_1\right]^{-1}\delta s^{+}({\bf x},{\bf p},0)\nonumber\\
		&+\frac{1}{2}\int d{\bf x}\int d{\bf p}\eta({\bf x})\left[\nu-\mathcal{L}_2\right]^{-1}\delta s^{-}({\bf x},{\bf p},0),\\
		D_{\parallel}(\nu)=&1-\frac{\Gamma_c}{4}\int d{\bf x}\int d{\bf p}\eta({\bf x})\left[\nu-\mathcal{L}_1\right]^{-1}\eta e^{-iK}\rho\nonumber\\
		&-\frac{\Gamma_c}{4}\int d{\bf x}\int d{\bf p}\eta({\bf x})\left[\nu-\mathcal{L}_2\right]^{-1}\eta e^{iK}\rho.
	\end{align}
	Using the actual form of the propagators
	\begin{align*}
		e^{\mathcal{L}_1t}g({\bf x})=&e^{-i\frac{\Gamma_c}{2}\int_{0}^t\eta\left({\bf x}-\frac{\bf p}{m}\tau\right)J^{\parallel}_0d\tau}g\left({\bf x}-\frac{\bf p}{m}t\right)\\
		=&e^{i\left[K\left({\bf x}-\frac{\bf p}{m}t,{\bf p}\right)-K\left({\bf x},{\bf p}\right)\right]}g\left({\bf x}-\frac{\bf p}{m}t\right),\\
		e^{\mathcal{L}_2t}g({\bf x})=&e^{i\frac{\Gamma_c}{2}\int_{0}^t\eta\left({\bf x}-\frac{\bf p}{m}\tau\right)J^{\parallel}_0d\tau}g\left({\bf x}-\frac{\bf p}{m}t\right)\\
		=&e^{i\left[K\left({\bf x},{\bf p}\right)-K\left({\bf x}-\frac{\bf p}{m}t,{\bf p}\right)\right]}g\left({\bf x}-\frac{\bf p}{m}t\right)
	\end{align*}
	and Eq.~\eqref{sz0}, we obtain the final result given in Eq.~\eqref{DispersionHiggs}.
	
	\section{\label{App:Goldstone}Stability of the SSR phase: Goldstone mode}
	In this section we show the details of the derivation for the Goldstone mode dispersion relation given by Eq.~\eqref{DispersionGoldstone2}.
	
	The stability of the Goldstone mode can be calculated by solving
	\begin{align*}
		\frac{\partial \delta s^{\perp}}{\partial t}+\frac{\bf p}{m}\cdot\nabla_{\bf x} \delta s^{\perp}=&\frac{\Gamma_c}{2}\eta({\bf x}) \delta J^{\perp}s^{z}_0({\bf x},{\bf p}).
	\end{align*}
	Laplace transformation leads to
	\begin{align}
		\left[\nu-\mathcal{L}_0\right]L[\delta s^{\perp}]=&\delta s^{\perp}({\bf x},{\bf p},0)+\frac{\Gamma_c}{2}L[\delta J^{\perp}]\eta({\bf x})s^{z}_0({\bf x},{\bf p}),
	\end{align}
	where we used the definition of Eq.~\eqref{L0}. Using the same steps as in Appendix~\ref{App:NSR} we find
	\begin{align}
		L[\delta J^{\perp}]=\frac{A^{\perp}(\nu)}{D_{\perp}(\nu)},
	\end{align}
	with
	\begin{align}
		A^{\perp}(\nu)=&\int d{\bf x}\int d{\bf p}\eta({\bf x})\left[\nu-\mathcal{L}_0\right]^{-1}\delta s^{\perp}({\bf x},{\bf p},0),\\
		D_{\perp}(\nu)=&1-\frac{\Gamma_c}{2}\int d{\bf x}\int d{\bf p}\eta({\bf x})\left[\nu-\mathcal{L}_0\right]^{-1}\eta s^{z}_0.
	\end{align}
	Using Eq.~\eqref{propagator} we find the result \begin{align}
		\label{DispersionGoldstone}
		D_{\perp}(\nu)=&1-\frac{\Gamma_c}{2}\int_{0}^{\infty} dt e^{-\nu t}\int d{\bf x}\int d{\bf p}\eta\left({\bf x}+\frac{\bf p}{m}t\right)\eta s^{z}_0.
	\end{align}
	
	This dispersion relation, just like the dispersion relation for the Higgs mode, simplifies to Eq.~\eqref{Dispersionrelationnonsupp} in the limit $J^{\parallel}_0\to0$. Let us emphasize that the dispersion relations for the Higgs and the Goldstone look very similar but are only equivalent in the NSR phase. In fact in the superradiant phase one main difference between the Higgs and Goldstone modes is that the latter is always undamped. This can be seen using Eq.~\eqref{Eqsparasteady} such that we can transform the dispersion relation~\eqref{DispersionGoldstone} to
	\begin{align*}
		D_{\perp}(\nu)=&1-\frac{\int_{0}^{\infty} dt e^{-\nu t}\int d{\bf x}\int d{\bf p}\eta\left({\bf x}+\frac{\bf p}{m}t\right)\frac{\bf p}{m}\cdot\nabla_{\bf x}s^\parallel_0}{J^\parallel_0}.
	\end{align*}
	For this and the following equations we use the notation $s^{\parallel}_0=s^{\parallel}_0({\bf x},{\bf p})$.
	Applying Gau\ss{} theorem and explicitly using the fact that the atoms enter in $\estate$ and that the mode function vanishes at infinity, we get
	\begin{align*}
		D_{\perp}(\nu)=&1+\frac{\int_{0}^{\infty} dte^{-\nu t}\int d{\bf x}\int d{\bf p}\frac{d}{dt}\eta\left({\bf x}+\frac{\bf p}{m}t\right)s^\parallel_0}{J^\parallel_0}.
	\end{align*}
	After another partial integration we obtain the final result visible in Eq.~\eqref{DispersionGoldstone2} where we have used Eq.~\eqref{Jparallel}.
	
	\section{\label{App:LinewidthNSR}Linewidth in the NSR phase}
	This section provides details of the calculations of the $g_1^x$ function in the NSR phase.
	
	In order to do this we integrate Eq.~\eqref{sxdensity} where we assume $s^z=\rho({\bf p})$ and drop second order terms in the noise contribution. This integration is done using the characteristics method. Defining $s^x(t)=s^x[{\bf x}_i+{\bf p}(t-t_i)/m,t]$, with ${\bf x}_i=(-x_i,y_i,z_i)$ the position where the atom enters the cavity and $t_i$ the initial time, we obtain
	\begin{align*}
		s^x(t)=&s^x(t_i)+\int_{t_i}^tdt'\,\eta\left[{\bf x}(t')\right]\left[\frac{\Gamma_c}{2}J^x(t')+\mathcal{F}^x(t')\right]\rho,
	\end{align*}
	where ${\bf x}(t')={\bf x}_i+{\bf p}(t'-t_i)/m$. We can now use ${t-t_i=m(x+x_i)/p_x}$ to express $s^x(t_i)=W^x(y_i,z_i,{\bf p},t_i)$
	where $y_i=y-p_y(x+x_i)/p_x$, $z_i=z-p_z(x+x_i)/p_x$, and $t_i=t-m(x+x_i)/p_x$. After a change of variables $t'\mapsto t-t'$ we get
	\begin{align*}
		s^x(t)=&s^x(t_i)\\&+\int_{0}^{\infty}dt'\,\eta\left[{\bf x}(t-t')\right]\left[\frac{\Gamma_c}{2}J^x(t-t')+\mathcal{F}^x(t-t')\right]\rho,
	\end{align*}
	where we extend the integral to infinity because we assume that $\eta({\bf x})=0$ for $x<-x_i$. Furthermore ${{\bf x}(t-t')={\bf x}-{\bf p}t'/m}$ is independent of $t$. Multiplying $s^x(t)$ by $\eta({\bf x})$ and integrating over the phase space leads to a linear equation for $J^x$. This can be solved using the Laplace transformation and we get 
	\begin{align}
		L[J^x]=&\frac{L[J_{W^x}]+2\frac{1-D(\nu)}{\Gamma_c}L\left[\mathcal{F}^{x}\right]}{D(\nu)},\label{LJX}
	\end{align}
	where $D(\nu)$ is the dispersion relation in Eq.~\eqref{Dispersionrelationnonsupp}, and
	\begin{align}
		\label{JW}
		J_{W^x}(t) = &\int d{\bf x}\int d{\bf p}\,\eta\left({\bf x}\right)W^{x}\left(y_i,z_i,{\bf p},t_i\right).
	\end{align}
	Notice that $y_i$ and $z_i$ depend on ${\bf x}$ and ${\bf p}$. The time $t_i$ depends on ${\bf x}$, ${\bf p}$, and $t$.
	Since we are in the NSR regime we expect all zeros of $D(\nu)$ to be negative. We denote now by $\nu_0$ the zero with the largest real part. We assume in the following that this is a zero of first order. In the long time limit we can conclude that, defining the inverse of the residue of $1/D(\nu)$ as
	\begin{align}
		C_0=\lim_{\nu\to \nu_0}\frac{D(\nu)}{\nu-\nu_0},
	\end{align}
	the dipole is given by
	\begin{align}
		J^x(t)\approx J_{\mathrm{in}}^x(t)+J_{\mathrm{c}}^x(t).
	\end{align}
	where
	\begin{align}
		J_{\mathrm{in}}^x(t)=&\frac{\int_{0}^{t} dt'\,e^{\nu_0 (t-t')}\int d{\bf x}\int d{\bf p}\,\eta\left({\bf x}\right)W^{x}\left(y_i,z_i,{\bf p},t_i'\right)}{C_0},\\
		J_{{\mathrm{c}}}^x(t)=&\frac{\int_{0}^{t} dt'\,e^{\nu_0 (t-t')}\frac{2}{\Gamma_c}\mathcal{F}^x(t')}{C_0},
	\end{align}
	originate from the noise introduced by the incoming atoms and by the cavity noise, respectively. Here, $t_i'=t'-m(x+x_i)/p_x$.
	
	Since the cavity noise and the input noise are independent, the $g_1^x$ function is now completely determined by
	\begin{align}
		g_1^x(t) = \langle J^x(t+t_0)J^x(t_0)\rangle\approx&\, g_{1,\mathrm{in}}^x(t)+g_{1,\mathrm{c}}^x(t),
	\end{align}
	where
	\begin{align}
		g_{1,\mathrm{in}}^x(t)=&\langle J_{\mathrm{in}}^x(t+t_0)J_{\mathrm{in}}^x(t_0)\rangle,\\
		g_{1,\mathrm{c}}^x(t)=&\langle J_{\mathrm{c}}^x(t+t_0)J_{\mathrm{c}}^x(t_0)\rangle.
	\end{align}
	It is straightforward to calculate the cavity noise that takes the form
	\begin{align}
		\label{cavitynoiseterm}
		g_{1,\mathrm{c}}^x(t)=\frac{2e^{\nu_0t}}{\nu_0 \Gamma_cC_0^2}.
	\end{align}
	For the calculations of the contribution of the incoming atoms we use the noise correlations that are defined in Eq.~\eqref{noise}.
	The input noise term takes the form
	\begin{align}
		g_{1,\mathrm{in}}^x(t)=&\frac{\int_{0}^{t+t_0} dt'\int_0^{t_0} dt''e^{\nu_0(t+2t_0-t'-t'')}\chi(t'-t'')}{C_0^2},
	\end{align}
	where
	\begin{align}
		\label{chi}
		\chi(t'-t'')=\int d{\bf x}\int d{\bf p}\rho({\bf p})\eta\left[{\bf x}+\frac{\bf p}{m}(t'-t'')\right]\eta\left({\bf x}\right).
	\end{align}
	While the actual form of this integral is dependent on the distribution and the mode function $\eta$, we can still analyze it in the limit where the time is much larger than the transit time $\tau$. For a time $t'\gg\tau$ we obtain $\eta\left({\bf x}+\frac{\bf p}{m}t'\right)\eta\left({\bf x}\right)\approx 0$. Therefore it is reasonable to define
	\begin{align}
		\label{tchar}
		t_{\mathrm{char}}=&\int_{-\infty}^{\infty} dt'\chi(t'),
	\end{align}
	and approximate
	\begin{align}
		\label{approx}
		\chi(t'-t'')\approx t_{\mathrm{char}}\delta(t'-t'').
	\end{align}
	Here $t_{\mathrm{char}}$ is the characteristic timescale for the decay of $\chi$. Using Eq.~\eqref{approx} we can calculate
	\begin{align}
		g_{1,\mathrm{in}}^x(t)\approx&\frac{t_{\mathrm{char}}e^{\nu_0t}}{2\nu_0C_0^2}.\label{inputnoiseterm}
	\end{align}
	We emphasize that the actual form of $g_{1,\mathrm{in}}^x(t)$ for small $t\lesssim\tau$ depends on the density $\rho({\bf p})$ and the mode function $\eta({\bf x})$. However, the results in Eq.~\eqref{cavitynoiseterm} and in Eq.~\eqref{inputnoiseterm} show that the long time behavior ($t\gg\tau$) of the $g_1$ function can be described by an exponential with decay $\nu_0$.
	
	\section{\label{App:LinewidthSSR}Linewidth in the SSR phase}
	In this section we show how we find the linewidth $\Gamma$ given by Eq.~\eqref{Gamma}. 
	
	We use Eq.~\eqref{orths} to calculate $s^{\perp}(t)$. Multiplying it by $\eta({\bf x})$ and integrating over the whole phase space, we obtain $J^{\perp}$. The resulting equation can be solved using a Laplace transformation where we eventually get
	\begin{align}
		\label{LJperp}
		L[J^{\perp}]\approx&\frac{L[J_{W^{\perp}}]+2\frac{1-D_{\perp}(\nu)}{\Gamma_c}L\left[\mathcal{F}^{\perp}\right]}{D_{\perp}(\nu)}.
	\end{align}
	This result is completely equivalent to Eq.~\eqref{LJX} except we use now the dispersion relation of the Goldstone mode in Eq.~\eqref{DispersionGoldstone2}. The noise equivalent to Eq.~\eqref{JW} is given by
	\begin{align}
		J_{W^{\perp}}(t)=&\int d{\bf x}\int d{\bf p}\,\eta\left({\bf x}\right)W^{\perp}\left(y_i,z_i,{\bf p},t_i\right).\label{JWperp}
	\end{align}
	The main difference between Eq.~\eqref{LJX} and Eq.~\eqref{LJperp} is the different zeros of the dispersion relations in Eq.~\eqref{Dispersionrelationnonsupp} and Eq.~\eqref{DispersionGoldstone}.  While the zero of Eq.~\eqref{Dispersionrelationnonsupp} always results in an exponential behavior, the dominant zero of Eq.~\eqref{DispersionGoldstone} is $\nu_0=0$. This implies that the dynamics of $J^{\perp}$ and the resulting phase $\varphi=J^{\perp}/J_0^{\parallel}$ are diffusive.
	
	For simplicity let us again assume that $\nu_0=0$ is a first order zero of Eq.~\eqref{DispersionGoldstone}. In that case we can define a non-vanishing
	\begin{align}
		C_{\perp}=&\lim_{\nu\to 0}\frac{D_{\perp}(\nu)}{\nu}\nonumber\\
		=&\frac{\int_{0}^{\infty} dt\int d{\bf x}\int d{\bf p}\eta\left({\bf x}+\frac{\bf p}{m}t\right)s^\parallel_0}{J^\parallel_0},\label{Cperp}
	\end{align}
	and use it to obtain
	\begin{align}
		J^{\perp}(t)\approx J_{\mathrm{in}}^{\perp}(t)+J_{\mathrm{c}}^{\perp}(t),
	\end{align}
	where
	\begin{align}
		J_{\mathrm{in}}^{\perp}(t)=&\frac{\int_{0}^{t} dt'\,\int d{\bf x}\int d{\bf p}\,\eta\left({\bf x}\right)W^{\perp}\left(y_i,z_i,{\bf p},t_i'\right)}{C_{\perp}},\\
		J_{{\mathrm{c}}}^{\perp}(t)=&\frac{\int_{0}^{t} dt'\,\frac{2}{\Gamma_c}\mathcal{F}^{\perp}(t')}{C_{\perp}},
	\end{align}
	are the input and cavity noise terms, respectively.
	
	We can now give a simple expression for the $g_1$ function
	\begin{align}
		g_1(t)\approx\lim_{t_0\to\infty}\frac{(J_0^{\parallel})^2}{4} e^{-\frac{\langle \Delta\varphi(t,t_0)^2 \rangle}{2}},
	\end{align}
	where $\Delta\varphi(t,t_0)=\varphi(t+t_0)-\varphi(t_0)$. Let us without loss of generality choose $t_0=0$ and write $\Delta\varphi(t,0)=\Delta\varphi(t)$. Since input noise and cavity noise are independent, we obtain
	\begin{align}
		\langle\Delta\varphi(t)^2\rangle=\langle\Delta\varphi_{\mathrm{in}}(t)^2\rangle+\langle\Delta\varphi_{\mathrm{c}}(t)^2\rangle,
	\end{align}
	with $\Delta\varphi_{\mathrm{in}}(t)=J_{\mathrm{in}}^{\perp}/J_0^{\parallel}$ and $\Delta\varphi_{\mathrm{c}}(t)=J_{\mathrm{c}}^{\perp}/J_0^{\parallel}$. 
	
	The term corresponding to the cavity noise is given by
	\begin{align}
		\langle\Delta\varphi_{\mathrm{c}}(t)^2\rangle=\frac{4}{\Gamma_cC_{\perp}^2(J_0^{\parallel})^2}t,
	\end{align}
	showing the usual increase of the variance with $t$ of a diffusion process.
	
	For the noise term that arises from incoming atoms, we use Eq.~\eqref{noise} to obtain
	\begin{align}
		\langle\Delta\varphi_{\mathrm{in}}(t)^2\rangle=&\frac{\int_{0}^{t} dt'\int_0^{t} dt''\chi(t'-t'')}{C_{\perp}^2(J_0^{\parallel})^2},
	\end{align}
	where we have used the definition in Eq.~\eqref{chi}. While this process has a non-trivial time dependence for $t\lesssim\tau$ we can write in the large time limit $t\gg\tau$ the following expression
	\begin{align}
		\langle\Delta\varphi_{\mathrm{in}}(t)^2\rangle\approx&\frac{t_{\mathrm{char}}}{C_{\perp}^2(J_0^{\parallel})^2}t,
	\end{align}
	with the characteristic timescale $t_{\mathrm{char}}$ defined in Eq.~\eqref{tchar}. In the long-time limit this leads to the result shown in Eq.~\eqref{g1dyn} and Eq.~\eqref{Gamma}.

	\section{Spontaneous emission and dephasing}\label{App:spontaneous}
	In this section we discuss how we can simulate spontaneous emission and dephasing. We also discuss when we can neglect these effects.
	
	In the description that we have used for the main part of the paper we have neglected free-space spontaneous emission with rate $\gamma_1$ as well as $T_2$ dephasing. This can be justified if $\gamma_1\tau \ll 1$ and $\tau/T_2\ll1$. In this limit, both effects are negligible during the transit time of an atom, and the corresponding noise is dominated by input noise and cavity shot noise. In order to observe superradiance we require $N\Gamma_c\tau>1$, which results in $N\Gamma_c\gg \gamma_1$ given $\gamma_1\tau\ll1$. This means that we assume a large collective cooperativity $N\mathcal{C}=Ng^2/(\kappa\gamma_1)\gg1$.
	
	We will now show how we can add the effects of spontaneous emission and dephasing to our model. For this we now generalize the master equation in Eq.~\eqref{Mastereq} to
	\begin{align}
		\label{Mastereq_spont}
		\frac{d \hat{\rho}}{d t}= \frac{1}{i\hbar}\left[\hat{H},\hat{\rho}\right]
		+ \kappa \lindblad[\hat{a}]\hat{\rho}
		+  \sum_j\left\{\gamma_1 \lindblad[\hat{\sigma}_j^-]
		+ \frac{\gamma_2}{4} \lindblad[\hat{\sigma}_j^z]\right\}\hat{\rho},
	\end{align}
	where $\gamma_2 = 2/T_2$ is the rescaled $T_2$ dephasing rate~\cite{Meiser:2009}. Using this master equation, we can eliminate the cavity field and derive the full $c$-number Heisenberg-Langevin equations. These $c$-number stochastic differential equations for the dipole components are given by
	\begin{align}
		\label{sx_spont}
		\frac{ds_j^{x}}{dt}=&\frac{\Gamma_c}{2}\eta({\bf x}_j)s_j^{z}J^{x}
		- \frac{\gamma_{1} + \gamma_{2} }{2}  s^x_j
		+  \mathcal{F}^{x}_{j},\\
		\label{sy_spont}
		\frac{ds_j^{y}}{dt}=&\frac{\Gamma_c}{2}\eta({\bf x}_j)s_j^{z}J^{y}
		- \frac{\gamma_{1} + \gamma_{2}}{2}  s^y_j
		+ \mathcal{F}^{y}_{j},\\
		\label{sz_spont}
		\frac{ds_j^{z}}{dt}=&-\frac{\Gamma_c}{2}\eta({\bf x}_j)\left(J^{x}s_j^{x}+J^{y}s_j^{y}\right)
		- \gamma_{1}(s_j^z + 1)
		+ \mathcal{F}^{z}_{j},
	\end{align}
	where we have used noise terms  $\mathcal{F}^\alpha_j=\mathcal{S}^\alpha_j+\mathcal{F}^{\alpha}_{j,\gamma_1}+\mathcal{F}^{\alpha}_{j,\gamma_2}$ for $\alpha\in\{x,y,z\}$. While the noise terms $\mathcal{S}^\alpha_j$ have been given in Eqs.~\eqref{sx}--\eqref{x}, we now introduce two additional independent noise sources $\mathcal{F}^{\alpha}_{j,\gamma_1}$ and $\mathcal{F}^{\alpha}_{j,\gamma_2}$, which originate from spontaneous emission and $T_2$ dephasing, respectively. These noise terms fulfill $\langle\mathcal{F}^{\alpha}_{j,\gamma_1}(t)\mathcal{F}^{\beta}_{k,\gamma_1}(t')\rangle=2\left(D_{j,\gamma_1}\right)_{\alpha\beta}\delta_{jk}\delta(t-t')$ and $\langle\mathcal{F}^{\alpha}_{j,\gamma_2}(t)\mathcal{F}^{\beta}_{k,\gamma_2}(t')\rangle=2\left(D_{j,\gamma_2}\right)_{\alpha\beta}\delta_{jk}\delta(t-t')$, with the diffusion matrices given by
	\begin{equation}
		D_{j,\gamma_1}
		=
		\kbordermatrix{
			& \beta=x & y & z \\
			\alpha=x & 1 & 0 & s_j^x\\
			y & 0 & 1 & s_j^y \\
			z & s_j^x & s_j^y & 2(1+s_j^z)\\}
		\times \frac{\gamma_1}{2}
	\end{equation}
	and
	\begin{equation}
		D_{j,\gamma_2}
		=
		\kbordermatrix{
			& \beta=x & y & z \\
			\alpha=x & 1 & 0 & 0\\
			y & 0 & 1 & 0 \\
			z & 0 & 0 & 0\\}
		\times \frac{\gamma_2}{2}.
	\end{equation}
	We simulate Eqs.~\eqref{sx_spont}--\eqref{sz_spont} for the numerical results we present in Sec.~\ref{subsec: spontaneous}.


\begin{thebibliography}{99}
		
		\bibitem{Baumann:2010}
		K. Baumann, C. Guerlin, F. Brennecke, and T. Esslinger, 
		Dicke quantum phase transition with a superfluid gas in an optical cavity, 
		Nature 
		{\bf 464}, 1301 (2010).
		
		\bibitem{Ritsch:2013}
		H. Ritsch, P. Domokos, F. Brennecke, and T. Esslinger, 
		Cold atoms in cavity-generated dynamical optical potentials, 
		Rev. Mod. Phys. 
		{\bf 85}, 553 (2013).
		
		\bibitem{Habibian:2013}
		H. Habibian, A. Winter, S. Paganelli, H. Rieger, and G. Morigi, 
		Bose-Glass Phases of Ultracold Atoms due to Cavity Backaction, 
		Phys. Rev. Lett. 
		{\bf110}, 075304 (2013).
		
		\bibitem{Vaidya:2018}
		V. D. Vaidya, Y. Guo, R. M. Kroeze, K. E. Ballantine, A. J. Koll\'ar, J. Keeling, and B. L. Lev, Tunable-Range, Photon-Mediated Atomic Interactions in Multimode Cavity QED, 
		Phys. Rev. X 
		{\bf 8}, 011002 (2018).
		
		\bibitem{Muniz:2020}
		J. A. Muniz, D. Barberena, R. J. Lewis-Swan, D. J. Young, J. R. K. Cline, A. M. Rey, and J. K. Thompson, 
		Exploring dynamical phase transitions with cold atoms in an optical cavity, 
		Nature 
		{\bf580}, (2020).
		
		\bibitem{Schaeffer:2017}
		S. A. Sch\"affer, B. T. R. Christensen, M. R. Henriksen, and J. W. Thomsen, Dynamics of bad-cavity-enhanced interaction with cold Sr atoms for laser stabilization, 
		Phys. Rev. A 
		{\bf 96}, 013847 (2017).
		
		\bibitem{Swan:2018}
		R. J. Lewis-Swan, M. A. Norcia, J. R. K. Cline, J. K. Thompson, and A. M. Rey, 
		Robust Spin Squeezing via Photon-Mediated Interactions on an Optical Clock Transition, 
		Phys. Rev. Lett. 
		{\bf 121}, 070403 (2018).
		
		\bibitem{Norcia:2018}
		M. A. Norcia, J. R. K. Cline, J. A. Muniz, J. M. Robinson, R. B. Hutson, A. Goban, G. E. Marti, J. Ye, and J. K. Thompson,
		Frequency Measurements of Superradiance from the Strontium Clock Transition, 
		Phys. Rev. X 
		{\bf 8}, 021036 (2018).
		
		\bibitem{Gothe:2019}
		H. Gothe, D. Sholokhov, A. Breunig, M. Steinel, and J. Eschner, 
		Continuous-wave virtual-state lasing from cold ytterbium atoms, 
		Phys. Rev. A 
		{\bf99}, 013415 (2019).
		
		\bibitem{Pedrozo:2020}
		E. Pedrozo-Pe\~n{}afiel, S. Colombo, C. Shu, A. F. Adiyatullin, Z. Li, E. Mendez, B. Braverman, A. Kawasaki, D. Akamatsu, Y. Xiao, and V. Vuleti\'c, 
		Entanglement on an optical atomic-clock transition, 
		Nature 
		{\bf 588}, 414 (2020).
		
		\bibitem{Dicke:1954}
		{R. H. Dicke}, 
		{Coherence in Spontaneous Radiation Processes}, 
		{Phys. Rev.} 
		{\bf 93}, 99 (1954).
		
		\bibitem{Gross:1982}
		{M. Gross and S. Haroche}, 
		{Superradiance: An essay on the theory of collective spontaneous emission}, 
		{Phys. Rep.} 
		{\bf 93}, 301 (1982).
		
		\bibitem{Meiser:2009}
		{D. Meiser, J. Ye, D. R. Carlson, and M. J. Holland}, 
		{Prospects for a Millihertz-Linewidth Laser}, 
		{Phys. Rev. Lett.} 
		{\bf 102}, 163601 (2009).
		
		\bibitem{Bohnet:2012}
		{J. G. Bohnet, Z. Chen, J. M. Weiner, D. Meiser, M. J. Holland, and J. K. Thompson}, 
		{A steady-state superradiant laser with less than one intracavity photon}, 
		{Nature (London)} 
		{\bf 484}, 78 (2012).
		
		\bibitem{Meiser:2010:1}
		{D. Meiser and M. J. Holland}, 
		{Steady-state superradiance with alkaline-earth-metal atoms}, 
		{Phys. Rev. A} 
		{\bf 81}, 033847 (2010).
		
		\bibitem{Norcia:2016:1}
		{M. A. Norcia and J. K. Thompson}, 
		{Cold-Strontium Laser in the Superradiant Crossover Regime}, 
		{Phys. Rev. X} 
		{\bf 6}, 011025 (2016).
		
		\bibitem{Norcia:2016:2}
		{M. A. Norcia, M. N. Winchester, J. R. K. Cline, and J. K. Thompson}, 
		{Superradiance on the millihertz linewidth strontium clock transition}, 
		{Sci. Adv.} 
		{\bf 2}, e1601231 (2016).
		
		\bibitem{Xu:2014}
		{M. Xu, D. A. Tieri, E. C. Fine, J. K. Thompson, and M. J. Holland}, 
		{Synchronization of Two Ensembles of Atoms}, 
		{Phys. Rev. Lett.} 
		{\bf 113}, 154101 (2014).
		
		\bibitem{Zhu:2015}
		{B. Zhu, J. Schachenmayer, M. Xu, F. Herrera, J. G. Restrepo, M. J. Holland, and A. M. Rey},
		{Synchronization of interacting quantum dipoles}, 
		{New J. Phys.} 
		{\bf 17}, 083063 (2015).
		
		\bibitem{Weiner:2017}
		{J. M. Weiner, K. C. Cox, J. G. Bohnet, and J. K. Thompson},
		{Phase synchronization inside a superradiant laser}, 
		{Phys. Rev. A} 
		{\bf 95}, 033808 (2017).
		
		\bibitem{Gong:2018}
		{Z. Gong, R. Hamazaki, and M. Ueda}, 
		{Discrete Time-Crystalline Order in Cavity and Circuit QED Systems}, 
		{Phys. Rev. Lett.} 
		{\bf 120}, 040404 (2018). 
		
		\bibitem{Iemini:2018}
		{F. Iemini, A. Russomanno, J. Keeling, M. Schir{\`o}, M. Dalmonte, and R. Fazio}, 
		{Boundary Time Crystals}, 
		{Phys. Rev. Lett.} 
		{\bf 121}, 035301 (2018). 
		
		\bibitem{Tucker:2018}
		{K. Tucker, B. Zhu, R. J. Lewis-Swan, J. Marino, F. Jimenez, J. G. Restrepo, and A. M. Rey}, 
		{Shattered time: can a dissipative time crystal survive many-body correlations?}, 
		{New J. Phys.} 
		{\bf 20}, 123003 (2018). 
		
		\bibitem{Barberena:2019}
		{D. Barberena, R. J. Lewis-Swan, J. K. Thompson, and A. M. Rey}, 
		{Driven-dissipative quantum dynamics in ultra-long-lived dipoles in an optical cavity}, 
		{Phys. Rev. A} 
		{\bf 99}, 053411 (2019). 
		
		\bibitem{Booker:2020}
		{C. Booker, B. Bu{\v{c}}a, and D. Jaksch}, 
		{Non-stationarity and dissipative time crystals: spectral properties and finite-size effects}, 
		{New J. Phys.} 
		{\bf 22}, 085007 (2020). 
		
		\bibitem{Kessler:2020}
		{H. Ke{\ss}ler, J. G. Cosme, C. Georges, L. Mathey, and A. Hemmerich}, 
		{From a continuous to a discrete time crystal in a dissipative atom-cavity system}, 
		{New J. Phys.} 
		{\bf 22}, 085002 (2020). 
		
		\bibitem{Kessler:2021}
		H. Ke\ss{}ler, P. Kongkhambut, C. Georges, L. Mathey, J. G. Cosme, A. Hemmerich, 
		Observation of a dissipative time crystal, 
		preprint arXiv:2012.08885 (2021).
		
		\bibitem{Chen:2009}
		{J. Chen}, 
		{Active optical clock}, 
		{Chin. Sci. Bull.} 
		{\bf 54}, 348 (2009).
		
		\bibitem{Zhang:2013}
		{T. Zhang, Y. Wang, X. Zang, W. Zhuang, and J. Chen}, 
		{Active optical clock based on four-level quantum system}, 
		{Chin. Sci. Bull.} 
		{\bf 58}, 2033 (2013).
		
		\bibitem{Meiser:2010:2}
		{D. Meiser and M. J. Holland}, 
		{Intensity fluctuations in steady-state superradiance}, 
		{Phys. Rev. A} 
		{\bf 81}, 063827 (2010).
		
		\bibitem{Maier:2014}
		{T. Maier, S. Kraemer, L. Ostermann, and H. Ritsch}, 
		{A superradiant clock laser on a magic wavelength optical lattice}, 
		{Opt. Express} 
		{\bf 22}, 13269 (2014).
		
		\bibitem{Kraemer:2016}
		{S. Kr{\"a}mer, L. Ostermann, and H. Ritsch}, 
		{Optimized geometries for future generation optical lattice clocks}, 
		{EPL} 
		{\bf 114}, 14003 (2016).   
		
		\bibitem{Debnath:2018}
		{K. Debnath, Y. Zhang, and K. M\o{}lmer}, 
		{Lasing in the superradiant crossover regime}, 
		{Phys. Rev. A} 
		{\bf 98}, 063837 (2018).  
		
		\bibitem{Zhang:2018}
		Y. Zhang, Y.-X. Zhang and K. M\o{}lmer, Monte-Carlo simulations of superradiant lasing, 
		New J. Phys. 
		{\bf 20}, 112001 (2018).
		
		\bibitem{Laske:2019}
		T. Laske, H. Winter, and A. Hemmerich, Pulse Delay Time Statistics in a Superradiant Laser with Calcium Atoms, Phys. Rev. Lett. {\bf123}, 103601 (2019).
		
		\bibitem{Schaeffer:2020}
		S. A. Sch\"affer, M. Tang, M. R. Henriksen, A. A. J\o{}rgensen, B. T. R. Christensen, and J. W. Thomsen, Lasing on a narrow transition in a cold thermal strontium ensemble,
		Phys. Rev. A {\bf 101}, 013819 (2020).
		
		\bibitem{Zhang:2021}
		Y. Zhang, C. Shan, and K. M\o{}lmer, Ultranarrow Superradiant Lasing by Dark Atom-Photon Dressed States, 
		Phys. Rev. Lett. 
		{\bf 126}, 123602 (2021).
		
		\bibitem{Liu:2020}
		{H. Liu, S. B. J{\"a}ger, X. Yu, S. Touzard, A. Shankar, M. J. Holland, and T. L. Nicholson}, 
		{Rugged mHz-Linewidth Superradiant Laser Driven by a Hot Atomic Beam}, 
		{Phys. Rev. Lett.} 
		{\bf 125}, 253602 (2020).  
		
		\bibitem{Jaeger:2021}
		{S. B. J{\"a}ger, H. Liu, A. Shankar, J. Cooper, and M. J. Holland}, 
		{Regular and bistable steady-state superradiant phases of an atomic beam traversing an optical cavity}, 
		{Phys. Rev. A} 
		{\bf 103}, 013720 (2021). 
		
		\bibitem{Temnov:2005}
		{V. V. Temnov}, 
		{Superradiance and subradiance in the overdamped many-atom micromaser}, 
		{Phys. Rev. A} 
		{\bf 71}, 053818 (2005).  
		
		\bibitem{Bonifacio:1971}
		{R. Bonifacio, P. Schwendimann, and F. Haake}, 
		{Quantum Statistical Theory of Superradiance. I}, 
		{Phys. Rev. A} 
		{\bf 4}, 302 (1971). 
		
		\bibitem{Schachenmayer:2015}
		{J. Schachenmayer, A. Pikovski, and A. M. Rey}, 
		{Many-Body Quantum Spin Dynamics with Monte Carlo Trajectories on a Discrete Phase Space}, 
		{Phys. Rev. X} 
		{\bf 5},  011022 (2015).
		
		\bibitem{DeGiorgio:1970}
		V. DeGiorgio and M. O. Scully, Analogy between the Laser Threshold Region and a Second-Order Phase Transition, 
		Phys. Rev. A 
		{\bf 2}, 1170 (1970).
		
		\bibitem{Higgs:1964}
		P. W. Higgs, Broken Symmetries and the Masses of Gauge Bosons, 
		Phys. Rev. Lett. 
		{\bf 13}, 508 (1964).
		
		\bibitem{Englert:1964}
		F. Englert and R. Brout, Broken Symmetry and the Mass of Gauge Vector Mesons, 
		Phys. Rev. Lett. 
		{\bf 13}, 321 (1964).
		
		\bibitem{Goldstone:1961}
		J. Goldstone, 
		Field theories with Superconductor solutions, 
		Nuovo Cim. 
		{\bf 19}, 154 (1961)
		
		\bibitem{Goldstone:1962}
		J. Goldstone, A. Salam, and S. Weinberg, 
		Broken Symmetries, 
		Phys. Rev. 
		{\bf127}, 965 (1962).
		
		\bibitem{Lamb:1999}
		W. E. Lamb, W. P. Schleich, M. O. Scully, and C. H. Townes, 
		Laser physics: Quantum controversy in action, 
		Rev. Mod. Phys. 
		{\bf 71}, (1999).
		
		\bibitem{Shankar:2021}
		{A. Shankar, J. T. Reilly, S. B. J\"ager, M. J. Holland}, 
		{Subradiant-to-Subradiant Phase Transition in the Bad Cavity Laser}, 
		{preprint arXiv:2103.07402} (2021).
		
		\bibitem{Bonifacio:1994a}
		R. Bonifacio and L. De Salvo, 
		Collective atomic recoil laser (CARL) optical gain without inversion by collective atomic recoil and self-bunching of two-level atoms, Nucl. Instrum. Methods, 
		Phys. Res., Sect. A 
		{\bf 341}, 360 (1994).
		
		\bibitem{Bonifacio:1994b}
		R. Bonifacio, L. De Salvo, L. M. Narducci, and E. J. D'Angelo, 
		Exponential gain and self-bunching in a collective atomic recoil laser, 
		Phys. Rev. A 
		{\bf 50}, 1716 (1994).    
		
		\bibitem{Bonifacio:2005}
		R. Bonifacio, M. M. Cola, N. Piovella and G. R. M. Robb, 
		A quantum model for collective recoil lasing, 
		Europhys. Lett. 
		{\bf69}, 55 (2005).
		
		\bibitem{Slama:2007}
		S. Slama, S. Bux, G. Krenz, C. Zimmermann, and Ph. W. Courteille, 
		Superradiant Rayleigh Scattering and Collective Atomic Recoil Lasing in a Ring Cavity, 
		Phys. Rev. Lett. 
		{\bf 98}, 053603 (2007).
		
		\bibitem{Jaeger:2019}
		{S. B. J{\"a}ger, J. Cooper, M. J. Holland, and G. Morigi}, 
		{Dynamical Phase Transitions to Optomechanical Superradiance}, 
		{Phys. Rev. Lett.} 
		{\bf 123}, 053601 (2019).
		
		\bibitem{Jaeger:2020}
		{S. B. J{\"a}ger, M. J. Holland, and G. Morigi}, 
		{Superradiant optomechanical phases of cold atomic gases in optical resonators}, 
		{Phys. Rev. A} 
		{\bf 101}, 023616 (2020).   
		
		\bibitem{Chen:2019}
		C.-C. Chen, S. Bennetts, R. Gonz\'alez Escudero, B. Pasquiou, and F. Schreck, 
		Continuous Guided Strontium Beam with High Phase-Space Density, 
		Phys. Rev. Applied 
		{\bf 12}, 044014 (2019).
		
	\end{thebibliography}
\end{document}